\documentclass[11pt]{article}

\usepackage[T1]{fontenc}
\usepackage[utf8]{inputenc}
\usepackage{adjustbox}
\usepackage[margin=1in]{geometry}
\usepackage{setspace}
 \usepackage{rotating}
\usepackage{amsmath}
\usepackage{amssymb}
\usepackage{mathtools}
\usepackage{amsthm}
\allowdisplaybreaks
\numberwithin{equation}{section}   

\theoremstyle{plain}
\newtheorem{theorem}{Theorem}[section]

\newtheorem{proposition}[theorem]{Proposition}

\theoremstyle{definition}

\theoremstyle{remark}

\DeclareMathOperator*{\argmax}{arg\,max}
\DeclareMathOperator*{\argmin}{arg\,min}

\usepackage{lmodern}

\usepackage{graphicx}
\usepackage{booktabs}
\usepackage{array}
\usepackage{multirow}
\usepackage{threeparttable}   
\usepackage{siunitx}          
\usepackage{rotating}

\usepackage{enumitem}         

\usepackage{microtype}
\usepackage{xcolor}

\usepackage{tikz}
\usepackage{subcaption}

\usetikzlibrary{angles,arrows.meta,backgrounds,calc,quotes}

\usepackage[round,authoryear]{natbib}
\usepackage[hidelinks]{hyperref}

\usepackage{algorithm}        
\usepackage{algpseudocode}    

\usepackage{longtable}
\usepackage{threeparttablex} 
\usepackage{booktabs}

\begin{document}

\begin{center}
{\LARGE\bf Distributionally Robust Recovery of Omitted Factors from Forecast Residuals with Application to Interest Rate Risk Management}
\end{center}

\vspace{2em}

\begin{singlespace}
\begin{center}
{\large Jinjun Liu}\\[2pt]
{\small Department of Mathematics, Hong Kong Baptist University, Hong Kong SAR}\\
{\small \texttt{22482865@life.hkbu.edu.hk}}

\vspace{0.9em}

{\large Ming-Yen Cheng}\\[2pt]
{\small Department of Mathematics, Hong Kong Baptist University, Hong Kong SAR}\\
{\small \texttt{chengmingyen@hkbu.edu.hk}}

\vspace{0.9em}

{\small \today}
\end{center}
\end{singlespace}

\begin{abstract}
A forecasting model compresses its predictors into an estimate of a conditional mean, and the
systematic structure that estimate omits survives in the second moment of its forecast errors.
Accuracy comparisons do not measure this structure, and variance-based extraction does not recover
the part of it that a given decision bears. In this paper, we propose a distributionally robust
framework that recovers the omitted structure from the residuals of a fixed forecaster: a decision
is made robust over a two-layer moment ambiguity set on the standardized residual cross-section,
and the discovery statistic is the covariance forcing, the component of the decision's residual
risk transverse to its exposure. We demonstrate that the forcing is invariant to shrinkage and to
isotropic inflation of the covariance, so the recovered direction is a property of the residuals
rather than of the regularization, the sense in which the recovery is ground truth.  This
is confirmed on monthly U.S.\ Treasury zero-coupon yields from 2006 to 2025, where the recovered
factor is named by factor-adjusted robust selection against a panel of $111$ macroeconomic,
Treasury supply-and-demand, and financial indicators. From the residuals of the linear
factor-augmented dynamic Nelson--Siegel benchmark the factor names as a leading business-cycle
factor, anchored on the Conference Board leading index and certified by a block-permutation test;
from those of the more accurate nonlinear random forest benchmark the same procedure selects the
same real-activity family without certification. A neutralization test completes the evidence:
removing the recovered factor from the deployed duration position leaves volatility essentially
unchanged and worsens the tail, so the factor is a material systematic risk the position bears.
\end{abstract}
\noindent\textbf{Keywords:} distributionally robust optimization; moment ambiguity set;
forecast residuals; omitted factors; yield curve; interest rate risk management

\section{Introduction}
\label{sec:intro}

Interest-rate risk management is a decision made on a forecast. A manager chooses a duration
exposure across maturities whose profit and loss is driven by the next month's yield changes, and a
forecast of those changes is the input on which the choice rests. Policy announcements,
central-bank communication, and headline-driven speculation move yields without an immediate
fundamental counterpart \citep{baker2016measuring}, so the object of value is not the most accurate
forecast but a decision robust to what the forecast misses.

What the forecast misses has a definite location. Any fixed forecaster, linear or nonlinear,
compresses its predictors into an estimate of the conditional mean, and the systematic structure it
omits survives in the second moment of its forecast errors. Accuracy metrics are silent on this
structure, and variance-based extraction returns the largest residual direction, which need not be
the one a given position bears. The omitted structure nevertheless re-enters the decision, because
the residual covariance is the risk model of every position built on the forecast.

In this paper, we recover that structure from the residuals of a fixed forecaster and return it to
the decision the forecast serves. A duration position is made robust over the two-layer moment
ambiguity set of \citet{delage2010distributionally} placed on the standardized residual
cross-section, and the recovery is read off the optimizer. With $\hat{\Sigma}_{z,t}$ the residual
covariance n the training set through $t$ and $a_t$ the volatility-scaled robust exposure, the discovery
statistic is the covariance forcing
\begin{equation}
  \hat{d}_t = P^{\perp}_{a_t}\,\hat{\Sigma}_{z,t}\,a_t,
  \qquad
  v_t = \hat{d}_t / \lVert \hat{d}_t \rVert_2 ,
\label{eq:forcing_intro}
\end{equation}
where $P^{\perp}_{a_t}$ projects onto the hyperplane orthogonal to $a_t$; Section~\ref{sec:framework}
makes each object precise. The paper's principal contribution is theoretical, in three results: the
forcing is invariant to shrinkage and to isotropic inflation of the covariance, so the recovered
direction is a property of the residuals rather than of the regularization, the sense in which the
recovery is ground truth; it identifies the loading of a single omitted factor up to sign and an
explicit angle bound whenever that loading is neither aligned with nor orthogonal to the exposure
and its signal dominates the residual anisotropy; and under several factors it is an
exposure-weighted mixture of their projected loadings. The two layers stand on opposite sides of the invariance: the covariance layer perturbs the second
moment by exactly the scalings and inflations the forcing cannot see, so robustness in that layer
alone is regularization; the least-favorable shift of the mean layer lies along the residual risk
$\hat{\Sigma}_{z,t}a_t$ itself, and its component the exposure does not span is the recovered
direction. The guarantees act on the residual covariance
alone, so they hold for linear and nonlinear forecasters alike, and recovery is sharpest on the
residuals of linear benchmarks, whose restrictive span leaves the omitted structure in place. A
factor-adjusted robust regression then names the factor against a panel of $111$ macroeconomic,
Treasury supply-and-demand, and financial indicators, and the same two-layer program robustifies the
duration decision into which the factor feeds back.

The framework draws on three literatures. Moment-constrained ambiguity sets receive their
data-driven formulation, with finite-sample confidence radii, in \citet{delage2010distributionally};
Wasserstein balls center the ambiguity on the empirical distribution
\citep{MohajerinEsfahani2018,kuhn2019wasserstein}; and residual-based formulations place it on the
errors of an estimated model
\citep{nguyen2020distributionally,KannanBayraksanLuedtke2024,NguyenZhangWangBlanchetDelageYe2024}.
Throughout, the robust program is the end product. Here it is also an instrument: the optimizer's
exposure indexes a statistic of the residual covariance, and that statistic carries the discovery.

Factor extraction from large panels proceeds by variance or by the first moment. Principal
components summarize a panel by its leading directions \citep{BaiNg2002,ahn2013eigenvalue}; weak
factors can escape a variance ranking \citep{Onatski2012}, motivating extractions tilted toward a
target \citep{LettauPelger2020}; and penalized regression selects indicators for a conditional mean
\citep{Tibshirani1996,fan2001variable} under design conditions a comoving panel violates
\citep{Wainwright2009}, which factor-adjusted selection restores \citep{fan2020factor}. Each device
targets the mean or the leading variance directions. The recovery here targets neither: it
extracts, from the residual second moment, the direction a fixed decision is exposed to, and names
it afterward.

The empirical setting is one-month-ahead forecasting of the U.S.\ Treasury curve. The
\citet{nelson1987parsimonious} cross-section, made dynamic by \citet{diebold2006forecasting} and
factor-augmented with macroeconomic information by \citet{fernandes2019dynamic}, defines the linear
benchmark, and the random forest of \citet{breiman2001random} relaxes the functional form, with the
caution that the gains from added complexity shrink when features are noisily observed
\citep{cartea2025limited}. Forecast combination
\citep{BatesGranger1969,yang2004combining,caldeira2016forecast} enters only as a downstream
baseline. The paper is decision-oriented recovery, complementary to accuracy-oriented work, and its
premise favors the restrictive benchmark: a linear model whose span had already absorbed the
omitted structure would leave nothing to recover.

The paper makes four contributions. First, and principally, it establishes the theory of
residual-based factor recovery, whose invariance, identification, and decomposition results make
the recovered direction a ground-truth property of the residual covariance, valid for any
forecaster whose residuals satisfy the stated conditions (Section~\ref{sec:framework}). Second, it
carries recovery to the residuals of a jointly fitted, multi-output random forest, so that the
residual cross-section of a nonparametric learner is a coherent recovery object
(Section~\ref{sec:benchmarks}). Third, on monthly U.S.\ Treasury zero-coupon yields from 2006 to
2025 it recovers and names a leading business-cycle factor from the linear benchmark's residuals,
and shows that the same procedure on the more accurate forest's residuals selects the same
real-activity family without certifying it (Sections~\ref{sec:naming}--\ref{sec:empirical}).
Fourth, it closes the loop into risk management: the same two-layer program supplies the robust
duration position, and a neutralization test on the deployed position shows that the recovered
factor had been offsetting its worst months over the sample, tail evidence that the direction is
systematic structure (Section~\ref{sec:dv01}).

The remainder of the paper is organized as follows. Section~\ref{sec:framework} develops the
recovery framework and its guarantees, Section~\ref{sec:benchmarks} specifies the two residual
generators, Section~\ref{sec:naming} constructs the naming procedure, Section~\ref{sec:empirical}
reports the empirical study, Section~\ref{sec:dv01} instantiates the duration decision and the
neutralization evidence, and Section~\ref{sec:conclusion} concludes. Pseudocode for every procedure in the paper is collected in Appendix~\ref{app:algorithms}.

\section{Residual-Based Distributionally Robust Factor Recovery}
\label{sec:framework}

This section develops the residual-based factor-recovery framework on the U.S.\ Treasury curve and
states the paper's theoretical results. We hold the forecasting model fixed, form a distributionally
robust decision over the standardized cross-section of its forecast errors
\citep{KannanBayraksanLuedtke2024,NguyenZhangWangBlanchetDelageYe2024}, and take a functional of the
optimal exposure, the covariance forcing, whose direction identifies the latent channel of a
potential omitted factor. What the forcing recovers depends on the residual covariance and not on
the model that produces it, so the method is model-generic; we instantiate it on two forecasters of
contrasting form, a linear factor-augmented dynamic Nelson--Siegel model (FADNS) and a nonlinear
random forest (Section~\ref{sec:benchmarks}). Section~\ref{subsec:setup} defines the residual
object, its latent channel, and its regularized covariance; Section~\ref{subsec:two_layer} the
two-layer moment ambiguity set and the robust exposure; and Section~\ref{subsec:forcing} the forcing
and the guarantees on the direction it recovers. The resulting factor is named in
Section~\ref{sec:naming}. 

\subsection{Residuals as the discovery object}
\label{subsec:setup}

Fix a forecasting model $\mathrm{b}$ for the U.S.\ Treasury curve. The model is used for one-month
prediction and duration risk management from past information alone---lagged yields together with a
monthly panel of macroeconomic series, Treasury supply-and-demand variables, and financial variables---and the decision it induces is made robust to driven by news headlines and speculation episodes
that move yields without a fundamental counterpart. We take the horizon to be one month, $h=1$, and do
not target longer horizons, because policy and geopolitical events shift the trend of the curve before
they register in the macroeconomic data, so that multi-step forecasts on a rolling window would chase
movements the predictors cannot yet see. Let $\hat{y}^{\mathrm{b}}_{t+h\mid t}(\tau)$ be the model's
point forecast for maturity $\tau$ made at origin $t$. The object of study is not the forecast but its
error at each maturity, stacked across the $N=15$ maturities,
\begin{equation}
  e^{\mathrm{b}}_{t+h}(\tau)=y_{t+h}(\tau)-\hat{y}^{\mathrm{b}}_{t+h\mid t}(\tau),
  \qquad
  e^{\mathrm{b}}_{t+h}=\big(e^{\mathrm{b}}_{t+h}(\tau_1),\ldots,e^{\mathrm{b}}_{t+h}(\tau_N)\big)^{\!\top}\in\mathbb{R}^{N}.
\label{eq:residual_vector}
\end{equation}

All moments are formed on an expanding training set containing every month up to the forecast origin
$t$; we write $M_t$ for its length, and deployment begins after the burn-in specified in
Section~\ref{sec:empirical}. The
maturities differ in yield volatility, so their co-movement is read only after a common scaling. Let
$\hat{\sigma}^{\mathrm{b}}_t(\tau)$ denote the standard deviation of
$e^{\mathrm{b}}_{\,\cdot}(\tau)$ and $\bar{e}^{\mathrm{b}}_t$ the mean of
$e^{\mathrm{b}}_{\,\cdot}$ over the window, and collect the maturity volatilities on the diagonal of
\begin{equation}
  D^{\mathrm{b}}_t=\operatorname{diag}\big(\hat{\sigma}^{\mathrm{b}}_t(\tau_1),\ldots,\hat{\sigma}^{\mathrm{b}}_t(\tau_N)\big)\in\mathbb{R}^{N\times N}.
\label{eq:vol_scaling}
\end{equation}
The standardized residual shock is then
\begin{equation}
  z^{\mathrm{b}}_{t+h}=\big(D^{\mathrm{b}}_t\big)^{-1}\big(e^{\mathrm{b}}_{t+h}-\bar{e}^{\mathrm{b}}_t\big)\in\mathbb{R}^{N}.
\label{eq:std_residual}
\end{equation}

The robust stage inverts and perturbs the second moment of these shocks, so we estimate it by
well-conditioned linear shrinkage. Let
$\hat{S}^{\mathrm{b}}_{z,t}=w^{-1}\sum_{s}z^{\mathrm{b}}_{s}(z^{\mathrm{b}}_{s})^{\top}$ be the sample
covariance of the standardized shocks over the training set and
$\bar{\mu}^{\mathrm{b}}_t=N^{-1}\operatorname{tr}\hat{S}^{\mathrm{b}}_{z,t}$ their average variance. The
estimator
\begin{equation}
  \hat{\Sigma}^{\mathrm{b}}_{z,t}=(1-\theta)\,\hat{S}^{\mathrm{b}}_{z,t}+\theta\,\bar{\mu}^{\mathrm{b}}_t\,I_N,
  \qquad \theta\in[0,1),
\label{eq:shrunk_cov}
\end{equation}
shrinks $\hat{S}^{\mathrm{b}}_{z,t}$ toward the scaled identity $\bar{\mu}^{\mathrm{b}}_t I_N$, where
$I_N$ is the $N\times N$ identity and the intensity $\theta$ is the data-driven optimal weight of
\citet{LedoitWolf2004} rather than a tuning choice. Regularization conditions the covariance but does
not by itself single out a residual direction, and the forcing is invariant to it
(Proposition~\ref{prop:shrinkage_invariance}), so the choice of $\theta$ does not bias the recovered
direction. A systematic factor the model omits would enter the second moment as a latent channel: a
unit loading $b_0\in\mathbb{R}^{N}$, $\lVert b_0\rVert_2=1$, along which the residual carries variance
$\sigma_f^{2}>0$, contributing a rank-one component $\sigma_f^{2}b_0 b_0^{\top}$ to
$\hat{\Sigma}^{\mathrm{b}}_{z,t}$ that is absent from the conditional-mean forecast. This rank-one
picture motivates and interprets the recovery; it is not a distributional assumption on the errors
(Section~\ref{subsec:forcing}). The loading $b_0$ is unobserved and need not lie among the high-variance
directions of $\hat{\Sigma}^{\mathrm{b}}_{z,t}$; recovering it is the task of the robust exposure of
Section~\ref{subsec:two_layer} and the covariance forcing of Section~\ref{subsec:forcing}.

\subsection{Two-layer moment ambiguity set and the robust exposure}
\label{subsec:two_layer}

The decision is a position $\omega\in\mathbb{R}^{N}$ across the $N$ maturities, read as a duration
exposure and given economic form in Section~\ref{sec:dv01}. Under the volatility scaling
$D^{\mathrm{b}}_t$ of \eqref{eq:vol_scaling} the position has volatility-scaled exposure
$a^{\mathrm{b}}_t=D^{\mathrm{b}}_t\omega$ and bears the standardized shock as
$(a^{\mathrm{b}}_t)^{\top}z^{\mathrm{b}}$, so its one-period return is
\begin{equation}
  r^{\mathrm{b}}_{t+h}(\omega)=\omega^{\top}\hat{\mu}^{\mathrm{b}}_t+(a^{\mathrm{b}}_t)^{\top}z^{\mathrm{b}}_{t+h},
\label{eq:position_return}
\end{equation}
where $\hat{\mu}^{\mathrm{b}}_t\in\mathbb{R}^{N}$ is the forecast-implied mean input across maturities,
specified in Section~\ref{sec:dv01}. The forecast enters the decision only through
$\hat{\mu}^{\mathrm{b}}_t$. The shock $z^{\mathrm{b}}_{t+h}$ is uncertain, and we make the position
robust over a two-layer moment ambiguity set \citep{delage2010distributionally}.

Write $z=z^{\mathrm{b}}_{t+h}$ for the standardized shock of \eqref{eq:std_residual} and let
$\mathbb{Q}$ denote a candidate distribution for it. The two-layer ambiguity set constrains the first
two moments of $\mathbb{Q}$,
\begin{equation}
  \mathcal{A}^{\mathrm{b}}_t(\rho)
  =\Big\{\,\mathbb{Q}\;:\;
    \mathbb{E}_{\mathbb{Q}}[z]^{\top}\big(\hat{\Sigma}^{\mathrm{b}}_{z,t}\big)^{-1}\mathbb{E}_{\mathbb{Q}}[z]\le\gamma_1,\;\;
    \mathbb{E}_{\mathbb{Q}}\big[z z^{\top}\big]\preceq\gamma_2\,\Sigma^{\mathrm{b}}_t(\rho)\Big\},
  \qquad
  \Sigma^{\mathrm{b}}_t(\rho)=\hat{\Sigma}^{\mathrm{b}}_{z,t}+\rho I_N ,
\label{eq:delage_ye_set}
\end{equation}
where $\hat{\Sigma}^{\mathrm{b}}_{z,t}$ is the shrunk standardized-shock covariance of
\eqref{eq:shrunk_cov} and $I_N$ is the identity on the $N$ maturities. The first constraint confines the
shock mean $\mathbb{E}_{\mathbb{Q}}[z]$ to an ellipsoid of shape $\hat{\Sigma}^{\mathrm{b}}_{z,t}$ and
size $\gamma_1\ge 0$ centered at the origin, the nominal mean of the de-meaned residual, and so absorbs
uncertainty in whether the residual is centered. The second constraint bounds the shock second moment
$\mathbb{E}_{\mathbb{Q}}[zz^{\top}]$ in the positive-semidefinite order by $\gamma_2\ge 1$ times the target
$\Sigma^{\mathrm{b}}_t(\rho)$, which inflates the estimated covariance isotropically by a radius
$\rho\ge 0$, with $\Sigma^{\mathrm{b}}_t(0)=\hat{\Sigma}^{\mathrm{b}}_{z,t}$, and so makes the risk robust
to the residual covariance exceeding its estimate.

Under a candidate $\mathbb{Q}$ the return \eqref{eq:position_return} has mean
$\omega^{\top}\hat{\mu}^{\mathrm{b}}_t+(a^{\mathrm{b}}_t)^{\top}\mathbb{E}_{\mathbb{Q}}[z]$ and
variance $(a^{\mathrm{b}}_t)^{\top}\operatorname{Cov}_{\mathbb{Q}}[z]\,a^{\mathrm{b}}_t$, the latter
controlled by the second-moment bound through
$\operatorname{Cov}_{\mathbb{Q}}[z]\preceq\mathbb{E}_{\mathbb{Q}}[zz^{\top}]\preceq\gamma_2\,\Sigma^{\mathrm{b}}_t(\rho)$.
The robust exposure maximizes the worst-case Sharpe ratio, its numerator the least favorable mean
over the set and its denominator the greatest standard deviation over the set,
\begin{equation}
  \omega^{\mathrm{b},\ast}_t(\rho)
  \in\argmax_{\omega\neq 0}\;
  \frac{\displaystyle\min_{\mathbb{Q}\in\mathcal{A}^{\mathrm{b}}_t(\rho)}\mathbb{E}_{\mathbb{Q}}\big[r^{\mathrm{b}}_{t+h}(\omega)\big]}
       {\displaystyle\max_{\mathbb{Q}\in\mathcal{A}^{\mathrm{b}}_t(\rho)}\big(\operatorname{Var}_{\mathbb{Q}}\big[r^{\mathrm{b}}_{t+h}(\omega)\big]\big)^{1/2}} .
\label{eq:robust_sharpe}
\end{equation}
The inner minimum and maximum are the worst-case mean and worst-case variance of a linear position under
first- and second-moment constraints; both are attained in closed form, reducing \eqref{eq:robust_sharpe}
to a deterministic program, which Proposition~\ref{prop:robust_reduction} solves.

\begin{proposition}[Reduction of the robust exposure]
\label{prop:robust_reduction}
Suppose $\hat{\Sigma}^{\mathrm{b}}_{z,t}\succ 0$ and $\gamma_1\le\gamma_2$, with
$a^{\mathrm{b}}_t=D^{\mathrm{b}}_t\omega$ the standardized exposure built from the volatility scaling
\eqref{eq:vol_scaling} and $\Sigma^{\mathrm{b}}_t(\rho)=\hat{\Sigma}^{\mathrm{b}}_{z,t}+\rho I_N$ the
inflated covariance of \eqref{eq:delage_ye_set}. For every $\omega$ with $a^{\mathrm{b}}_t\neq 0$, the
worst-case mean and worst-case variance of the position return \eqref{eq:position_return} over
$\mathcal{A}^{\mathrm{b}}_t(\rho)$ are
\begin{equation}
  \min_{\mathbb{Q}\in\mathcal{A}^{\mathrm{b}}_t(\rho)}\mathbb{E}_{\mathbb{Q}}\big[r^{\mathrm{b}}_{t+h}(\omega)\big]
  =\omega^{\top}\hat{\mu}^{\mathrm{b}}_t-\sqrt{\gamma_1}\,\big\lVert a^{\mathrm{b}}_t\big\rVert_{\hat{\Sigma}^{\mathrm{b}}_{z,t}},
  \qquad
  \max_{\mathbb{Q}\in\mathcal{A}^{\mathrm{b}}_t(\rho)}\operatorname{Var}_{\mathbb{Q}}\big[r^{\mathrm{b}}_{t+h}(\omega)\big]
  =\gamma_2\,\big\lVert a^{\mathrm{b}}_t\big\rVert^{2}_{\Sigma^{\mathrm{b}}_t(\rho)},
\label{eq:worst_case_moments}
\end{equation}
where $\lVert x\rVert_M=(x^{\top}Mx)^{1/2}$ denotes the norm induced by a positive-definite matrix $M$.
Consequently \eqref{eq:robust_sharpe} is equivalent to the deterministic fractional program
\begin{equation}
  \omega^{\mathrm{b},\ast}_t(\rho)\in\argmax_{\omega\neq 0}\;
  \frac{\omega^{\top}\hat{\mu}^{\mathrm{b}}_t-\sqrt{\gamma_1}\,\big\lVert a^{\mathrm{b}}_t\big\rVert_{\hat{\Sigma}^{\mathrm{b}}_{z,t}}}
       {\sqrt{\gamma_2}\,\big\lVert a^{\mathrm{b}}_t\big\rVert_{\Sigma^{\mathrm{b}}_t(\rho)}},
  \qquad a^{\mathrm{b}}_t=D^{\mathrm{b}}_t\omega .
\label{eq:robust_sharpe_reduced}
\end{equation}
\end{proposition}

\begin{proof}
The return $r^{\mathrm{b}}_{t+h}(\omega)=\omega^{\top}\hat{\mu}^{\mathrm{b}}_t+(a^{\mathrm{b}}_t)^{\top}z$
depends on the distribution $\mathbb{Q}$ only through the first two moments of the shock $z$, namely its
mean $m=\mathbb{E}_{\mathbb{Q}}[z]$ and second moment $M=\mathbb{E}_{\mathbb{Q}}[zz^{\top}]$, since
$\mathbb{E}_{\mathbb{Q}}[r^{\mathrm{b}}_{t+h}(\omega)]=\omega^{\top}\hat{\mu}^{\mathrm{b}}_t+(a^{\mathrm{b}}_t)^{\top}m$
and $\operatorname{Var}_{\mathbb{Q}}[r^{\mathrm{b}}_{t+h}(\omega)]=(a^{\mathrm{b}}_t)^{\top}(M-mm^{\top})a^{\mathrm{b}}_t$.
Each extremum may therefore be taken over the moment pairs $(m,M)$ admitted by
$\mathcal{A}^{\mathrm{b}}_t(\rho)$, which by \eqref{eq:delage_ye_set} are those with
$m^{\top}(\hat{\Sigma}^{\mathrm{b}}_{z,t})^{-1}m\le\gamma_1$ and $M\preceq\gamma_2\Sigma^{\mathrm{b}}_t(\rho)$
in the positive-semidefinite order. We treat the two extrema separately, as the objective
\eqref{eq:robust_sharpe} requires.

\emph{Worst-case mean.} This involves only $m$, through the linear term $(a^{\mathrm{b}}_t)^{\top}m$. Since
$\hat{\Sigma}^{\mathrm{b}}_{z,t}\succ 0$, the map $(x,y)\mapsto x^{\top}(\hat{\Sigma}^{\mathrm{b}}_{z,t})^{-1}y$
is an inner product, and the Cauchy--Schwarz inequality in that inner product gives, for every $m$ in the
mean ellipsoid,
\[
  (a^{\mathrm{b}}_t)^{\top}m
  =\big(\hat{\Sigma}^{\mathrm{b}}_{z,t}a^{\mathrm{b}}_t\big)^{\top}(\hat{\Sigma}^{\mathrm{b}}_{z,t})^{-1}m
  \ge-\big\lVert a^{\mathrm{b}}_t\big\rVert_{\hat{\Sigma}^{\mathrm{b}}_{z,t}}
        \big(m^{\top}(\hat{\Sigma}^{\mathrm{b}}_{z,t})^{-1}m\big)^{1/2}
  \ge-\sqrt{\gamma_1}\,\big\lVert a^{\mathrm{b}}_t\big\rVert_{\hat{\Sigma}^{\mathrm{b}}_{z,t}} .
\]
The lower bound is attained at
$m^{\star}=-\sqrt{\gamma_1}\,\hat{\Sigma}^{\mathrm{b}}_{z,t}a^{\mathrm{b}}_t/\lVert a^{\mathrm{b}}_t\rVert_{\hat{\Sigma}^{\mathrm{b}}_{z,t}}$,
which satisfies $m^{\star\top}(\hat{\Sigma}^{\mathrm{b}}_{z,t})^{-1}m^{\star}=\gamma_1$ and so lies on the
boundary of the ellipsoid. The point mass $\mathbb{Q}^{\star}=\delta_{m^{\star}}$ realizes this mean and
has second moment $M^{\star}=m^{\star}m^{\star\top}$. Because $m^{\star}m^{\star\top}$ is rank one,
$m^{\star}m^{\star\top}\preceq\gamma_1\hat{\Sigma}^{\mathrm{b}}_{z,t}$ is equivalent to
$m^{\star\top}(\gamma_1\hat{\Sigma}^{\mathrm{b}}_{z,t})^{-1}m^{\star}\le 1$, which holds with equality; hence
$M^{\star}\preceq\gamma_1\hat{\Sigma}^{\mathrm{b}}_{z,t}\preceq\gamma_1\Sigma^{\mathrm{b}}_t(\rho)\preceq\gamma_2\Sigma^{\mathrm{b}}_t(\rho)$,
using $\rho I_N\succeq 0$ and $\gamma_1\le\gamma_2$. Thus $\mathbb{Q}^{\star}\in\mathcal{A}^{\mathrm{b}}_t(\rho)$
and the bound is achieved, giving the first equality of \eqref{eq:worst_case_moments}.

\emph{Worst-case variance.} From $\operatorname{Var}_{\mathbb{Q}}[r^{\mathrm{b}}_{t+h}(\omega)]=(a^{\mathrm{b}}_t)^{\top}Ma^{\mathrm{b}}_t-((a^{\mathrm{b}}_t)^{\top}m)^{2}$,
the second-moment bound $M\preceq\gamma_2\Sigma^{\mathrm{b}}_t(\rho)$ gives
$(a^{\mathrm{b}}_t)^{\top}Ma^{\mathrm{b}}_t\le\gamma_2\,(a^{\mathrm{b}}_t)^{\top}\Sigma^{\mathrm{b}}_t(\rho)a^{\mathrm{b}}_t$,
while $-((a^{\mathrm{b}}_t)^{\top}m)^{2}\le 0$, so
$\operatorname{Var}_{\mathbb{Q}}[r^{\mathrm{b}}_{t+h}(\omega)]\le\gamma_2\lVert a^{\mathrm{b}}_t\rVert^{2}_{\Sigma^{\mathrm{b}}_t(\rho)}$.
Both inequalities are equalities at any distribution with $m=0$ and $M=\gamma_2\Sigma^{\mathrm{b}}_t(\rho)$,
for instance the Gaussian $\mathcal{N}(0,\gamma_2\Sigma^{\mathrm{b}}_t(\rho))$; this distribution satisfies
$0^{\top}(\hat{\Sigma}^{\mathrm{b}}_{z,t})^{-1}0=0\le\gamma_1$ and
$\gamma_2\Sigma^{\mathrm{b}}_t(\rho)\preceq\gamma_2\Sigma^{\mathrm{b}}_t(\rho)$, hence lies in
$\mathcal{A}^{\mathrm{b}}_t(\rho)$, and its variance equals the bound. This is the second equality of
\eqref{eq:worst_case_moments}.

Substituting the two values into \eqref{eq:robust_sharpe} yields \eqref{eq:robust_sharpe_reduced}. Each
inner extremum is attained by a distribution in $\mathcal{A}^{\mathrm{b}}_t(\rho)$, so the reduction is
exact and \eqref{eq:robust_sharpe} is a genuine minimax rather than a penalized surrogate.
\end{proof}

Program \eqref{eq:robust_sharpe_reduced} is invariant to the scale of $\omega$, so it is solved in
normalized form at every origin. At $\gamma_1=0$ the optimizer is available in closed form,
$a^{\mathrm{b}}_t\propto\big(\Sigma^{\mathrm{b}}_t(\rho)\big)^{-1}(D^{\mathrm{b}}_t)^{-1}\hat{\mu}^{\mathrm{b}}_t$,
which at $\rho=0$ is $(\hat{\Sigma}^{\mathrm{b}}_{z,t})^{-1}(D^{\mathrm{b}}_t)^{-1}\hat{\mu}^{\mathrm{b}}_t$.
For $\gamma_1>0$, fixing the worst-case standard deviation at one reduces
\eqref{eq:robust_sharpe_reduced} to the maximization of the robust numerator
$\omega^{\top}\hat{\mu}^{\mathrm{b}}_t-\sqrt{\gamma_1}\,\lVert a^{\mathrm{b}}_t\rVert_{\hat{\Sigma}^{\mathrm{b}}_{z,t}}$
subject to $\gamma_2\lVert a^{\mathrm{b}}_t\rVert^{2}_{\Sigma^{\mathrm{b}}_t(\rho)}\le 1$, a concave
program with a single convex quadratic constraint, solved numerically; a nonpositive optimum means the
standardized signal does not clear the mean penalty, and the position is withdrawn for that month. The
baseline exposure that indexes the recovery is the closed-form case,
$a^{\mathrm{b}}_t=D^{\mathrm{b}}_t\omega^{\mathrm{b},\ast}_t(0)$ at $\gamma_1=0$ and $\rho=0$; the
robust cases $\gamma_1>0$ and $\rho>0$ enter the risk decision of Section~\ref{sec:dv01}. With $N=15$
maturities each program is small, and the expanding training set keeps the reference covariance well
estimated.

\subsection{Covariance forcing and the recovered direction}
\label{subsec:forcing}

The discovery statistic is the covariance forcing. Applying the residual covariance to the baseline
exposure produces the \emph{risk vector}
\begin{equation}
  g^{\mathrm{b}}_t := \hat{\Sigma}^{\mathrm{b}}_{z,t}\,a^{\mathrm{b}}_t \in\mathbb{R}^{N},
\label{eq:risk_vector}
\end{equation}
the direction in which the position's residual risk acts. It splits into a component along the exposure
$a^{\mathrm{b}}_t$, which only rescales the position, and a component in the orthogonal hyperplane
$a_t^{\perp}:=\{x\in\mathbb{R}^{N}: (a^{\mathrm{b}}_t)^{\top}x=0\}$. The covariance forcing is the second
component, obtained by projecting $g^{\mathrm{b}}_t$ onto $a_t^{\perp}$,
\begin{equation}
  \hat{d}^{\mathrm{b}}_t=P^{\perp}_{a^{\mathrm{b}}_t}\,g^{\mathrm{b}}_t,
  \qquad
  P^{\perp}_{a}=I_N-\frac{a\,a^{\top}}{a^{\top}a},
  \qquad
  v^{\mathrm{b}}_t=\frac{\hat{d}^{\mathrm{b}}_t}{\lVert\hat{d}^{\mathrm{b}}_t\rVert_2},
\label{eq:forcing}
\end{equation}
where $P^{\perp}_{a^{\mathrm{b}}_t}$ is the orthogonal projector onto $a_t^{\perp}$ and $v^{\mathrm{b}}_t$
is the recovered direction. Geometrically, $\hat{d}^{\mathrm{b}}_t$ is the component of the risk vector
$g^{\mathrm{b}}_t$ transverse to the exposure, of length
$\lVert\hat{d}^{\mathrm{b}}_t\rVert_2=\lVert g^{\mathrm{b}}_t\rVert_2\sin\angle(g^{\mathrm{b}}_t,a^{\mathrm{b}}_t)$:
it is zero exactly when $g^{\mathrm{b}}_t$ is parallel to $a^{\mathrm{b}}_t$, that is, when
$a^{\mathrm{b}}_t$ is an eigenvector of $\hat{\Sigma}^{\mathrm{b}}_{z,t}$, and otherwise $v^{\mathrm{b}}_t$
points along the direction in $a_t^{\perp}$ that the position is subject to through $g^{\mathrm{b}}_t$
but does not span. The statistic is therefore decision-indexed: it depends on the exposure
$a^{\mathrm{b}}_t$, not on the leading eigenvector of $\hat{\Sigma}^{\mathrm{b}}_{z,t}$.

What the forcing recovers is determined by the residual covariance. The next three results all follow
from the identity $P^{\perp}_{a^{\mathrm{b}}_t}a^{\mathrm{b}}_t=0$ and hold for any covariance input and
baseline exposure, hence for the moment-based exposure of \eqref{eq:robust_sharpe}. The first
establishes that the forcing is invariant to the shrinkage of \eqref{eq:shrunk_cov}, because it sees
the covariance only through its anisotropic part, so the recovered direction does not depend on the
regularization; the remaining two identify that direction (Figure~\ref{fig:recovery_geometry}).

\begin{figure}[H]
\centering
\begin{subfigure}{0.48\textwidth}
\centering
\begin{tikzpicture}[>=Stealth, scale=0.92, line join=round]
  \draw[->, gray!70] (-3.0,0) -- (3.2,0) node[right] {\scriptsize $z_1$};
  \draw[->, gray!70] (0,-2.7) -- (0,3.0) node[above] {\scriptsize $z_2$};
  \fill (0,0) circle (1.0pt) node[below left=-2pt] {\scriptsize $0$};

  \begin{scope}[rotate around={30:(0,0)}]
    \draw[very thick, blue!60!black, fill=blue!10] (0,0) ellipse [x radius=2.35, y radius=1.85];
    \draw[thick, blue!55!black]                    (0,0) ellipse [x radius=1.75, y radius=1.25];
    \draw[thick, blue!55!black, dashed]            (0,0) ellipse [x radius=1.45, y radius=0.82];
  \end{scope}

  \begin{scope}[rotate around={30:(0,0)}]
    \draw[very thick, red!70!black, fill=red!14] (0,0) ellipse [x radius=0.68, y radius=0.39];
  \end{scope}
  \coordinate (M) at (0.32,0.28);
  \draw[->, red!70!black] (0,0) -- (M);
  \fill[red!70!black] (M) circle (1.3pt);
  \node[red!70!black] at (M) [above right=-3pt] {\scriptsize $\mathbb{E}_{\mathbb{Q}}[z]$};

  \node[red!70!black]  at (-1.55,2.35) {\scriptsize mean: $m^{\top}\hat{\Sigma}^{-1}_{z,t}m\le\gamma_1$};
  \draw[red!55,->] (-1.55,2.15) .. controls (-0.9,1.15) .. (-0.3,0.28);
  \node[blue!60!black] at (2.05,-2.05) {\scriptsize $\preceq\gamma_2\,\Sigma^{\mathrm{b}}_t(\rho)$};
  \draw[blue!45,->] (2.05,-1.85) .. controls (1.9,-1.2) .. (1.55,-0.85);
  \node[black!70] at (-1.95,-2.35) {\scriptsize dashed: $\hat{\Sigma}^{\mathrm{b}}_{z,t}$};
  \node[black!70] at (-1.55,-2.75) {\scriptsize solid: $+\rho I_N$};
\end{tikzpicture}
\caption{The ambiguity set $\mathcal{A}^{\mathrm{b}}_t(\rho)$ in the shock space: the mean layer (red)
bounds $\mathbb{E}_{\mathbb{Q}}[z]$, the covariance layer (blue) bounds $\mathbb{E}_{\mathbb{Q}}[zz^{\top}]$.}
\label{fig:ambiguity}
\end{subfigure}
\hfill
\begin{subfigure}{0.48\textwidth}
\centering
\begin{tikzpicture}[>=Stealth, scale=0.92, line join=round]
  \coordinate (O) at (0,0);
  \coordinate (A) at (3.7,0.85);      
  \coordinate (G) at (2.5,3.05);      
  \coordinate (P) at (2.6884,0.6173); 

  \draw[gray!55, dashed] ($(O)-0.15*(A)$) -- ($1.12*(A)$);
  \node[gray!70] at ($1.12*(A)$) [right=-2pt] {\scriptsize span $a^{\mathrm{b}}_t$};
  \draw[gray!55, dashed] ($(O)+0.40*(-0.85,3.7)$) -- ($(O)+0.24*(0.85,-3.7)$);
  \node[gray!70] at ($(O)+0.40*(-0.85,3.7)$) [above left=-3pt] {\scriptsize $a_t^{\perp}$};

  \draw[very thick, black] (O) -- (A) node[below right=-2pt] {$a^{\mathrm{b}}_t$};
  \draw[very thick, blue!65!black] (O) -- (G)
        node[above=-1pt] {\scriptsize $g^{\mathrm{b}}_t=\hat{\Sigma}^{\mathrm{b}}_{z,t}a^{\mathrm{b}}_t$};
  \draw[thick, black!55] (O) -- (P);
  \draw[very thick, red!75!black] (P) -- (G)
        node[midway, right=1pt] {\scriptsize $\hat{d}^{\mathrm{b}}_t$};
  \draw[dotted, black!50] (G) -- (P);
  \coordinate (Adir) at ($(P)+0.30*(A)$);
  \path pic[draw, angle radius=7pt] {right angle = Adir--P--G};
  \draw[very thick, red!75!black, ->] (O) -- ($(O)+0.28*(-0.85,3.7)$)
        node[left=1pt] {\scriptsize $v^{\mathrm{b}}_t$};
  \path pic[draw=blue!55!black, angle radius=13pt] {angle = A--O--G};
  \fill (O) circle (1.2pt);
\end{tikzpicture}
\caption{The forcing $\hat{d}^{\mathrm{b}}_t$ is the component of the risk vector $g^{\mathrm{b}}_t$
orthogonal to the exposure $a^{\mathrm{b}}_t$; the recovered direction $v^{\mathrm{b}}_t$ is its
normalization.}
\label{fig:forcing}
\end{subfigure}
\caption{Recovery geometry. Panel~(a) shows the two-layer moment ambiguity set
$\mathcal{A}^{\mathrm{b}}_t(\rho)$ of \eqref{eq:delage_ye_set}. The mean layer confines the shock mean
$\mathbb{E}_{\mathbb{Q}}[z]$ to the ellipsoid $\{m:m^{\top}(\hat{\Sigma}^{\mathrm{b}}_{z,t})^{-1}m\le\gamma_1\}$,
and the covariance layer bounds the second moment by $\gamma_2\,\Sigma^{\mathrm{b}}_t(\rho)$, the
reference covariance $\hat{\Sigma}^{\mathrm{b}}_{z,t}$ (dashed) after isotropic inflation to
$\Sigma^{\mathrm{b}}_t(\rho)=\hat{\Sigma}^{\mathrm{b}}_{z,t}+\rho I_N$ (solid) and scaling by
$\gamma_2\ge1$ (outer). Panel~(b) shows the forcing at the resulting robust exposure. The residual
covariance maps $a^{\mathrm{b}}_t$ to the risk vector
$g^{\mathrm{b}}_t=\hat{\Sigma}^{\mathrm{b}}_{z,t}a^{\mathrm{b}}_t$, which decomposes into a radial
component along $a^{\mathrm{b}}_t$ (grey), carrying no new direction, and the covariance forcing
$\hat{d}^{\mathrm{b}}_t=P^{\perp}_{a^{\mathrm{b}}_t}g^{\mathrm{b}}_t$ (red) in the orthogonal hyperplane
$a_t^{\perp}$. The recovered direction is $v^{\mathrm{b}}_t=\hat{d}^{\mathrm{b}}_t/\lVert\hat{d}^{\mathrm{b}}_t\rVert_2$,
and $\hat{d}^{\mathrm{b}}_t=0$ precisely when $a^{\mathrm{b}}_t$ is an eigenvector of
$\hat{\Sigma}^{\mathrm{b}}_{z,t}$.}
\label{fig:recovery_geometry}
\end{figure}

\begin{proposition}[Invariance to isotropic regularization]
\label{prop:shrinkage_invariance}
Let $a^{\mathrm{b}}_t\neq 0$ be any baseline exposure, let $S$ be any symmetric $N\times N$ matrix, and
write $\hat{d}(M)=P^{\perp}_{a^{\mathrm{b}}_t}\big(M a^{\mathrm{b}}_t\big)$ for the forcing built from a
matrix $M$. For every $\alpha>0$ and $\beta\in\mathbb{R}$,
\begin{equation}
  \hat{d}\big(\alpha S+\beta I_N\big)=\alpha\,\hat{d}(S),
\label{eq:forcing_invariance}
\end{equation}
so whenever $\hat{d}(S)\neq0$ the recovered direction is unchanged, including sign. In particular, for
the shrinkage estimator \eqref{eq:shrunk_cov} the choice $S=\hat{S}^{\mathrm{b}}_{z,t}$,
$\alpha=1-\theta>0$, $\beta=\theta\,\bar{\mu}^{\mathrm{b}}_t\ge0$ gives
$v^{\mathrm{b}}_t\big(\hat{\Sigma}^{\mathrm{b}}_{z,t}\big)=v^{\mathrm{b}}_t\big(\hat{S}^{\mathrm{b}}_{z,t}\big)$
for every intensity $\theta\in[0,1)$, and the isotropic inflation
$\Sigma^{\mathrm{b}}_t(\rho)=\hat{\Sigma}^{\mathrm{b}}_{z,t}+\rho I_N$ of \eqref{eq:delage_ye_set} is the
case $\alpha=1$, $\beta=\rho$, which likewise leaves the direction unchanged.
\end{proposition}

\begin{proof}
Write $a\equiv a^{\mathrm{b}}_t$. Since $P^{\perp}_{a}a=0$ and $P^{\perp}_{a}$ is linear,
\[
  \hat{d}\big(\alpha S+\beta I_N\big)
  =P^{\perp}_{a}\big((\alpha S+\beta I_N)a\big)
  =\alpha\,P^{\perp}_{a}\big(S a\big)+\beta\,P^{\perp}_{a}a
  =\alpha\,P^{\perp}_{a}\big(S a\big)
  =\alpha\,\hat{d}(S),
\]
which is \eqref{eq:forcing_invariance}. Because $\alpha>0$, normalization preserves both the vector and
its sign. The two stated cases substitute $(\alpha,\beta)=(1-\theta,\theta\bar{\mu}^{\mathrm{b}}_t)$ and
$(\alpha,\beta)=(1,\rho)$.
\end{proof}
The two layers of the ambiguity set stand on opposite sides of this invariance. The worst case of
the covariance layer is attained at the second moment
$\gamma_2\Sigma^{\mathrm{b}}_t(\rho)=\gamma_2\hat{\Sigma}^{\mathrm{b}}_{z,t}+\gamma_2\rho\,I_N$, the
case $(\alpha,\beta)=(\gamma_2,\gamma_2\rho)$ of \eqref{eq:forcing_invariance}, and at $\gamma_1=0$
the optimizer of \eqref{eq:robust_sharpe_reduced} is the closed form with
$\Sigma^{\mathrm{b}}_t(\rho)$ in place of $\hat{\Sigma}^{\mathrm{b}}_{z,t}$: robustness in the
second moment alone is the nominal decision under a regularized covariance, and by
Proposition~\ref{prop:shrinkage_invariance} the forcing cannot tell the two apart. The worst case
of the mean layer is the shift
$m^{\star}=-\sqrt{\gamma_1}\,\hat{\Sigma}^{\mathrm{b}}_{z,t}a^{\mathrm{b}}_t/\lVert a^{\mathrm{b}}_t\rVert_{\hat{\Sigma}^{\mathrm{b}}_{z,t}}$
attained in the proof of Proposition~\ref{prop:robust_reduction}, a move along the risk vector
\eqref{eq:risk_vector}, and projecting it off the exposure gives
$P^{\perp}_{a^{\mathrm{b}}_t}m^{\star}=-\sqrt{\gamma_1}\,\hat{d}^{\mathrm{b}}_t/\lVert a^{\mathrm{b}}_t\rVert_{\hat{\Sigma}^{\mathrm{b}}_{z,t}}$:
the component of the least-favorable mean the position does not span is the forcing, up to a
negative scale. For the recovery, whose baseline exposure is the closed-form case, the covariance
layer is therefore absorbed entirely by the invariance, while the mean layer's worst case realizes
itself along the very direction the statistic recovers.
The forcing statistic isolates the omitted structure exactly when the residual covariance departs
from isotropy through a single direction, and degrades gracefully when it does not.

\begin{proposition}[Single-factor recovery]
\label{prop:single_factor}
Suppose the standardized-shock covariance on the window admits the decomposition
\[
  \hat{\Sigma}^{\mathrm{b}}_{z,t}=\sigma_0^{2}I_N+\sigma_f^{2}b_0 b_0^{\top}+E,
\]
in which $\sigma_0^{2}\ge0$ is an isotropic background variance, the unit vector $b_0$, with
$\lVert b_0\rVert_2=1$, is the loading of a single residual factor carrying variance $\sigma_f^{2}>0$,
and the symmetric matrix $E=E^{\top}$ collects whatever anisotropy the first two terms leave
unexplained. Then the forcing statistic separates the omitted loading from that anisotropy,
\begin{equation}
  \hat{d}^{\mathrm{b}}_t=\sigma_f^{2}\,(b_0^{\top}a^{\mathrm{b}}_t)\,P^{\perp}_{a^{\mathrm{b}}_t}b_0
  +P^{\perp}_{a^{\mathrm{b}}_t}E\,a^{\mathrm{b}}_t,
\label{eq:single_factor_recovery}
\end{equation}
and two regimes follow. In the exact case, in which the anisotropy vanishes and the omitted loading is
neither orthogonal to nor aligned with the exposure, that is $E=0$ with $b_0^{\top}a^{\mathrm{b}}_t\neq0$
and $P^{\perp}_{a^{\mathrm{b}}_t}b_0\neq0$, the recovered direction is the omitted loading projected off
the exposure, up to sign,
\[
  v^{\mathrm{b}}_t=\operatorname{sign}(b_0^{\top}a^{\mathrm{b}}_t)\,
  P^{\perp}_{a^{\mathrm{b}}_t}b_0/\lVert P^{\perp}_{a^{\mathrm{b}}_t}b_0\rVert_2 .
\]
In the general case, let $s=\sigma_f^{2}\lvert b_0^{\top}a^{\mathrm{b}}_t\rvert\,
\lVert P^{\perp}_{a^{\mathrm{b}}_t}b_0\rVert_2$ measure the signal carried by the loading and
$\epsilon=\lVert P^{\perp}_{a^{\mathrm{b}}_t}E\,a^{\mathrm{b}}_t\rVert_2$ the contamination carried by the
anisotropy. Whenever the signal dominates, $s>\epsilon$, the forcing statistic is nonzero and its
deviation from the projected loading is controlled by their ratio,
\[
  \sin\angle\big(\hat{d}^{\mathrm{b}}_t,P^{\perp}_{a^{\mathrm{b}}_t}b_0\big)\le\epsilon/(s-\epsilon).
\]
\end{proposition}

\begin{proof}
Write $a$ for $a^{\mathrm{b}}_t$ throughout. Multiplying the decomposition on the right by $a$ gives
$\hat{\Sigma}^{\mathrm{b}}_{z,t}a=\sigma_0^{2}a+\sigma_f^{2}(b_0^{\top}a)b_0+Ea$. Because the projector
$P^{\perp}_{a}$ annihilates $a$, the isotropic term does not survive the projection, and applying
$P^{\perp}_{a}$ to both sides produces \eqref{eq:single_factor_recovery}. In the exact case the
anisotropic term is absent, so that $\hat{d}^{\mathrm{b}}_t=\sigma_f^{2}(b_0^{\top}a)P^{\perp}_{a}b_0$;
since $\sigma_f^{2}>0$ and $b_0^{\top}a\neq0$, normalizing this vector returns the direction
$v^{\mathrm{b}}_t$ displayed in the statement, with its sign inherited from $b_0^{\top}a$.

For the general case, decompose the forcing statistic as $\hat{d}^{\mathrm{b}}_t=u+\delta$, in which
$u=\sigma_f^{2}(b_0^{\top}a)P^{\perp}_{a}b_0$ carries the omitted loading and $\delta=P^{\perp}_{a}Ea$ is
the anisotropic contamination, so that $\lVert u\rVert_2=s$ and $\lVert\delta\rVert_2=\epsilon$ by
definition. When $s>\epsilon$ the triangle inequality gives
$\lVert\hat{d}^{\mathrm{b}}_t\rVert_2\ge\lVert u\rVert_2-\lVert\delta\rVert_2=s-\epsilon>0$, and the
statistic therefore does not vanish. Projecting onto the orthogonal complement of $u$ removes the signal
and leaves only the contamination, $P^{\perp}_{u}\hat{d}^{\mathrm{b}}_t=P^{\perp}_{u}\delta$, so that
\[
  \sin\angle\big(\hat{d}^{\mathrm{b}}_t,u\big)
  =\frac{\lVert P^{\perp}_{u}\hat{d}^{\mathrm{b}}_t\rVert_2}{\lVert\hat{d}^{\mathrm{b}}_t\rVert_2}
  =\frac{\lVert P^{\perp}_{u}\delta\rVert_2}{\lVert\hat{d}^{\mathrm{b}}_t\rVert_2}
  \le\frac{\epsilon}{s-\epsilon}.
\]
Because $u$ is a nonzero scalar multiple of $P^{\perp}_{a}b_0$ and the sine of an angle is insensitive to
the sign of either argument, the same bound governs $\angle(\hat{d}^{\mathrm{b}}_t,P^{\perp}_{a}b_0)$,
which is the assertion.
\end{proof}

The single-factor identity is the special case $K=1$, $\Sigma_u=\sigma_0^{2}I_N$ of a decomposition that
holds for any finite number of residual factors, though only the exact-recovery reading of the first
regime survives the generalization.

\begin{proposition}[Multi-factor decomposition]
\label{prop:multifactor}
Suppose the standardized-shock covariance admits the decomposition
$\hat{\Sigma}^{\mathrm{b}}_{z,t}=\Sigma_u+\sum_{k=1}^{K}\lambda_k b_k b_k^{\top}$, in which
$\Sigma_u\succeq0$ is an idiosyncratic covariance and each of the $K$ residual factors contributes a
loading $b_k\in\mathbb{R}^{N}$ weighted by its variance $\lambda_k\ge0$. Then the forcing statistic is the
superposition
\begin{equation}
  \hat{d}^{\mathrm{b}}_t=P^{\perp}_{a^{\mathrm{b}}_t}\Sigma_u a^{\mathrm{b}}_t
  +\sum_{k=1}^{K}\lambda_k\,(b_k^{\top}a^{\mathrm{b}}_t)\,P^{\perp}_{a^{\mathrm{b}}_t}b_k ,
\label{eq:multifactor_recovery}
\end{equation}
in which each factor enters through its own loading, projected off the exposure and weighted by both its
variance and its alignment $b_k^{\top}a^{\mathrm{b}}_t$ with that exposure, while the idiosyncratic
covariance contributes the single bias term $P^{\perp}_{a^{\mathrm{b}}_t}\Sigma_u a^{\mathrm{b}}_t$.
\end{proposition}

\begin{proof}
Write $a$ for $a^{\mathrm{b}}_t$. Because $b_k b_k^{\top}a=(b_k^{\top}a)b_k$, multiplying the decomposition
on the right by $a$ gives $\hat{\Sigma}^{\mathrm{b}}_{z,t}a=\Sigma_u a+\sum_{k=1}^{K}\lambda_k(b_k^{\top}a)b_k$.
Applying the linear projector $P^{\perp}_{a}$ term by term yields \eqref{eq:multifactor_recovery}.
\end{proof}

Proposition~\ref{prop:single_factor} shows that when a single residual direction dominates a weakly
anisotropic background, the forcing points along its non-radial part, up to sign and to an angle
governed by the ratio of anisotropy to signal; Proposition~\ref{prop:multifactor} shows that when
several directions are present, the forcing is an exposure-weighted mixture of their projected
loadings. Proposition~\ref{prop:shrinkage_invariance} carries both guarantees through the
regularization: under $\alpha\hat{\Sigma}^{\mathrm{b}}_{z,t}+\beta I_N$ the decomposition
\eqref{eq:multifactor_recovery} is unchanged in direction, and in the single-factor regime the
signal $s$ and the anisotropy $\epsilon$ both scale by $\alpha$, so the angle bound
$\sin\angle\le\epsilon/(s-\epsilon)$ is itself invariant. Because neither $v^{\mathrm{b}}_t$ nor its
bound depends on the shrinkage intensity $\theta$ or the inflation radius $\rho$, the recovered
direction is a property of the residuals themselves rather than an artifact of regularization, the
sense in which we call the recovery ground truth.

\section{Benchmark Forecasters as Residual Generators}
\label{sec:benchmarks}

The recovery framework of Section~\ref{sec:framework} operates on the residuals of a fixed
forecaster. We use two forecasters of contrasting form. The first is a factor-augmented dynamic
Nelson--Siegel model (FADNS), a parametric forecaster linear in its state. The second is a random
forest (RF), a nonparametric ensemble that admits nonlinear and interaction effects. The distinction
that matters for recovery is linear against nonlinear: the linear model leaves any nonlinear structure
in its residual, whereas the forest leaves in its residual only what a flexible nonlinear fit fails to
resolve. Both produce yield forecasts $\hat{y}^{\mathrm{b}}_{t+h\mid t}(\tau)$ on a rolling window of
$w=60$ months, indexed by $\mathrm{b}\in\{\mathrm{L},\mathrm{RF}\}$ with $\mathrm{L}$ denoting FADNS
and $\mathrm{RF}$ the forest, and the forecast error \eqref{eq:residual_vector} is the object the
framework recovers from. We report the one-month horizon $h=1$ throughout and retain the subscript
$t+h$. This section defines the two forecasts and their residuals; estimation and
forecast-evaluation detail are in Appendix~\ref{app:benchmarks}. The forest is run under ten independent seeds; the forecast carried into the recovery is the
seed average, so the residual \eqref{eq:resid_rf} is that of the seed-averaged forest, and the
accuracy comparison of Section~\ref{sec:empirical} reports the ten seeds individually as a range. A set of forecast-combination
schemes, used only as a downstream baseline in Section~\ref{subsec:emp_baselines}, is deferred to
Appendix~\ref{app:combination} and plays no part in recovery.

\subsection{Linear benchmark: factor-augmented dynamic Nelson--Siegel (FADNS)}
\label{subsec:fadns_family}

The dynamic Nelson--Siegel model \citep{diebold2006forecasting} writes the yield at maturity $\tau$ as
a linear combination of three latent factors,
\begin{equation}
  y_t(\tau)=\beta_{1t}+\beta_{2t}\,\frac{1-e^{-\lambda\tau}}{\lambda\tau}
  +\beta_{3t}\Big(\frac{1-e^{-\lambda\tau}}{\lambda\tau}-e^{-\lambda\tau}\Big)+u_t(\tau).
\label{eq:dns}
\end{equation}
These loadings make $\beta_{1t}$, $\beta_{2t}$, and $\beta_{3t}$ the level, slope, and curvature of
the curve; the decay parameter is fixed at $\lambda=0.0609$, and $u_t(\tau)$ is a cross-sectional
measurement error. Collecting the loadings across the $N=15$ maturities into
$\Lambda\in\mathbb{R}^{N\times 3}$, the factor vector
$\beta_t=(\beta_{1t},\beta_{2t},\beta_{3t})^{\top}$ is fitted at each $t$ by cross-sectional least
squares, $\hat{\beta}_t=(\Lambda^{\top}\Lambda)^{-1}\Lambda^{\top}y_t$, in which $y_t\in\mathbb{R}^{N}$
stacks the observed yields.

FADNS augments these factors with a lagged panel of macroeconomic, Treasury supply-and-demand, and
financial indicators \citep{fernandes2019dynamic}. The panel $Z_t\in\mathbb{R}^{p}$ collects $p=111$
monthly indicators in these three groups, each entering lagged by one month for real-time
availability; its composition is described in Section~\ref{sec:data} and listed in full in
Table~\ref{tab:EC_us_indicators} in Appendix~\ref{app:tables}. The leading $k$ principal components
of the panel, computed within the window, form the panel-factor vector $F^{(k)}_t\in\mathbb{R}^{k}$,
and the augmented state
$X^{(k)}_t=(\beta_{1t},\beta_{2t},\beta_{3t},F^{(k)\top}_t)^{\top}\in\mathbb{R}^{3+k}$ follows a
first-order vector autoregression,
\begin{equation}
  X^{(k)}_{t+1}=c^{(k)}+\Phi^{(k)}X^{(k)}_t+\eta^{(k)}_t,
\label{eq:fadns_var}
\end{equation}
with intercept $c^{(k)}$, coefficient matrix $\Phi^{(k)}$, and innovation $\eta^{(k)}_t$, all estimated
on the window. At the one-month horizon the factor forecast is the one-step conditional mean
$\hat{X}^{(k)}_{t+h\mid t}=c^{(k)}+\Phi^{(k)}X^{(k)}_t$, and its first three coordinates give
$\hat{\beta}_{t+h\mid t}$. Substituting into \eqref{eq:dns} yields the forecast
$\hat{y}^{\mathrm{L}}_{t+h\mid t}(\tau)$ and the residual
\begin{equation}
  e^{\mathrm{L}}_{t+h}(\tau)=y_{t+h}(\tau)-\hat{y}^{\mathrm{L}}_{t+h\mid t}(\tau).
\label{eq:resid_fadns}
\end{equation}
Plain DNS is the case $k=0$, in which $X^{(k)}_t=\beta_t$. We do not commit to a single factor count:
the empirical study carries the ten specifications $k=1,\ldots,10$ throughout
(Section~\ref{sec:empirical}), with estimation detail in Appendix~\ref{app:benchmarks}. Because the
FADNS forecast is linear in the level, slope, curvature, and leading panel components,
$e^{\mathrm{L}}$ retains what that linear span omits, namely weak, nonlinear, or interaction
structure, which is what the recovery targets.

\subsection{Nonlinear benchmark: random forest}
\label{subsec:rf_family}

The random forest \citep{breiman2001random} is a nonparametric benchmark that imposes no term-structure
form. It estimates the conditional mean of the yield vector by a data-driven partition of the predictor
space, the nonlinear counterpart of the linear span of FADNS. We fit a single \emph{joint} forest over
the entire maturity vector rather than one forest per maturity, so that its residual is a coherent
cross-maturity object, as the covariance forcing of \eqref{eq:forcing} requires. The forest estimates
\begin{equation}
  y_{t+h}=g_h(W_t)+\varepsilon_{t+h},
  \qquad g_h:[0,1]^{d_W}\to\mathbb{R}^{N},
  \qquad \mathbb{E}\!\left[\varepsilon_{t+h}\mid W_t\right]=0,
\label{eq:rf}
\end{equation}
in which $y_{t+h}\in\mathbb{R}^{N}$ is the yield vector, $g_h$ is the conditional-mean function,
$\varepsilon_{t+h}$ is the error term, and $W_t\in[0,1]^{d_W}$ is the scaled predictor defined next.

The raw predictor stacks the lagged indicators and the current and lagged yields,
\begin{equation}
  \check{W}_t=\big((Z_{t-\ell})_{\ell\in\mathcal{L}_Z},\,(y_{t-\ell})_{\ell\in\mathcal{L}_y}\big)\in\mathbb{R}^{d_W},
  \qquad d_W=|\mathcal{L}_Z|\,p+|\mathcal{L}_y|\,N,
\label{eq:rf_predictor}
\end{equation}
with lag sets $\mathcal{L}_Z=\{1,\ldots,60\}$ for the indicators, whose minimum one-month lag matches
their release calendar, and $\mathcal{L}_y=\{0,\ldots,59\}$ for the yields dated at origin $t$ and
earlier, so no forecast draws on contemporaneously unavailable data. Each coordinate is min--max
scaled to the unit interval over the estimation window $\mathcal{I}_t=\{t-w+1,\ldots,t\}$. Writing
$\underline{W}_{t,i}=\min_{s\in\mathcal{I}_t}\check{W}_{s,i}$ and
$\overline{W}_{t,i}=\max_{s\in\mathcal{I}_t}\check{W}_{s,i}$ for the smallest and largest values of
the $i$th coordinate across the window, the scaled coordinate is
$W_{t,i}=(\check{W}_{t,i}-\underline{W}_{t,i})/(\overline{W}_{t,i}-\underline{W}_{t,i})\in[0,1]$, and
the same affine map is applied to every in-window predictor, so the forest is fitted on the unit
cube. The response $y_{t+h}$ enters in its original units, so forecasts need no inverse transform.

The forest is the average of $B$ regression trees,
\begin{equation}
  g_h=\frac{1}{B}\sum_{b=1}^{B} g_h^{(b)},
\label{eq:rf_ensemble}
\end{equation}
each tree $g_h^{(b)}$ grown on the window by splitting to reduce the total squared error across the
$N$ maturities. Writing $[g_h(W_t)]_\tau=\hat{y}^{\mathrm{RF}}_{t+h\mid t}(\tau)$ for the forecast at
maturity $\tau$, the residual is
\begin{equation}
  e^{\mathrm{RF}}_{t+h}(\tau)=y_{t+h}(\tau)-\big[g_h(W_t)\big]_\tau.
\label{eq:resid_rf}
\end{equation}
Growing the forest jointly ties the per-maturity residuals through a shared partition, so
$e^{\mathrm{RF}}$ is comparable across maturities and its cross-section carries the structure the
forest does not resolve on the window. The tree-growing rule, the bootstrap resampling and feature
subsampling that decorrelate the trees, and the choice of $B$ and the remaining hyperparameters are
given in Appendix~\ref{app:benchmarks}.

Either benchmark's residuals now feed the recovery of Section~\ref{sec:framework}. Projecting the
realized standardized shock onto the direction recovered at the same origin gives, for each
$\mathrm{b}\in\{\mathrm{L},\mathrm{RF}\}$, the scalar series
\begin{equation}
  \xi^{\mathrm{b}}_{t+h}=\big(v^{\mathrm{b}}_{t}\big)^{\!\top} z^{\mathrm{b}}_{t+h},
\label{eq:projected_factor}
\end{equation}
in which $v^{\mathrm{b}}_{t}$ is recovered from information through origin $t$ and
$z^{\mathrm{b}}_{t+h}$ is the shock realized at $t+h$, so the series is formed out of sample. It
realizes the omitted factor over time but carries no economic label, having been extracted from the
residual second moment alone; it returns to the duration-risk decision in Section~\ref{sec:dv01},
and its economic content is supplied next, in Section~\ref{sec:naming}.

\section{Factor Naming via FarmSelect}
\label{sec:naming}

The recovery of Section~\ref{sec:framework} returns, for each benchmark $\mathrm{b}$ and origin $t$, the
direction $v^{\mathrm{b}}_t\in\mathbb{R}^{N}$ and the realized factor
$\xi^{\mathrm{b}}_{t+h}=(v^{\mathrm{b}}_t)^{\top}z^{\mathrm{b}}_{t+h}$ of \eqref{eq:projected_factor}, a
scalar series identified through the geometry of residual space alone and carrying no economic label.
This section supplies one by asking which observable indicators move with it.

The natural device is a regression of $\xi^{\mathrm{b}}_{t+h}$ on the panel of $p=111$ indicators, with
the nonzero coefficients read as the factor's economic content. Two properties of the panel defeat an
off-the-shelf penalized regression. The indicators comove strongly: the leading principal component
accounts for roughly $23\%$ of the panel's total variance (Table~\ref{tab:cumulative_variance} in
Appendix~\ref{app:tables}) and the first several components for a large majority, so that many indicators
are nearly collinear. A penalized fit to such a design distributes weight almost arbitrarily within each
collinear cluster, leaving the identity of the selected indicator unstable across resamples. The target
is also heavy-tailed, spiking on months dominated by a single news headline or policy announcement, so
that a squared-error fit would let those months govern the selection. We therefore use factor-adjusted
regularized model selection \citep{fan2020factor}, which separates the panel's pervasive comovement from
its variable-specific variation and penalizes only the latter, paired with a robust loss for the heavy
tails. The construction proceeds in three steps---extract the common factors, decorrelate, and
select---and throughout it acts on the covariance geometry of the indicator panel, the observable-space
analogue of the residual covariance on which the forcing statistic \eqref{eq:forcing} acts.

Recovery furnishes the naming sample one observation at a time. From the training set ending at month
$t$, the forcing statistic \eqref{eq:forcing} returns the direction $v^{\mathrm{b}}_t$, whose projection
gives the realized factor $\xi^{\mathrm{b}}_{t+h}$ of \eqref{eq:projected_factor}, paired with the
indicator vector $x_{t+h}$. We call each such month $t$ a \emph{naming origin} and write $n$ for their
number, so the naming sample is the $n$ pairs $\{(\xi^{\mathrm{b}}_{t+h},\,x_{t+h})\}$.

Write each standardized indicator vector as a common component plus an idiosyncratic remainder,
\begin{equation}
  x_{t+h}=B f_{t+h}+u_{t+h},
\label{eq:panel_factor}
\end{equation}
in which $x_{t+h}\in\mathbb{R}^{p}$ collects the $p=111$ indicators of Section~\ref{sec:benchmarks} dated
$t+h$ and standardized within the sample to zero mean and unit variance, $B\in\mathbb{R}^{p\times r}$ is
the loading matrix, $f_{t+h}\in\mathbb{R}^{r}$ holds the $r$ common factors driving the whole panel, and
$u_{t+h}\in\mathbb{R}^{p}$ is the idiosyncratic remainder, the part of each indicator not explained by
those factors. The common factors are the collinearity to be removed before selection, and the
idiosyncratic remainder carries the variable-specific, nameable signal; the three steps estimate this
decomposition and then select within the remainder.

The first step extracts the common factors. Stacking the $x_{t+h}^{\top}$ over the $n$ naming origins
into $X\in\mathbb{R}^{n\times p}$, we form the sample correlation matrix
$\hat{\Sigma}_X=n^{-1}X^{\top}X$ and its spectral decomposition
$\hat{\Sigma}_X=\sum_{j=1}^{p}\hat{\lambda}_j\hat{q}_j\hat{q}_j^{\top}$, with eigenvalues
$\hat{\lambda}_1\ge\cdots\ge\hat{\lambda}_p\ge0$ and orthonormal eigenvectors $\hat{q}_j$. The number of
common factors is set by the eigenvalue-ratio rule \citep{ahn2013eigenvalue},
\[
  \hat{r}=\argmax_{1\le j\le r_{\max}}\frac{\hat{\lambda}_j}{\hat{\lambda}_{j+1}},
\]
which cuts where consecutive eigenvalues drop most sharply, the bound being fixed at $r_{\max}=8$. The
estimated loadings and factor scores are then
\[
  \hat{B}=\big[\sqrt{\hat{\lambda}_1}\,\hat{q}_1,\ldots,\sqrt{\hat{\lambda}_{\hat{r}}}\,\hat{q}_{\hat{r}}\big]
  \in\mathbb{R}^{p\times\hat{r}},
  \qquad
  \hat{f}_{t+h}=(\hat{B}^{\top}\hat{B})^{-1}\hat{B}^{\top}x_{t+h}\in\mathbb{R}^{\hat{r}},
\]
so that $\hat{f}_{t+h}$ holds the leading $\hat{r}$ principal-component scores of $x_{t+h}$ normalized to
unit variance and column $j$ of $\hat{B}$ is the panel's loading on the $j$th factor.

The second step decorrelates. Subtracting the fitted common component from each indicator leaves the
estimated idiosyncratic remainder
\begin{equation}
  \hat{u}_{t+h}=x_{t+h}-\hat{B}\hat{f}_{t+h}\in\mathbb{R}^{p},
\label{eq:decorrelated}
\end{equation}
the empirical counterpart of $u_{t+h}$ in \eqref{eq:panel_factor}. With the strong common factors
projected out, the coordinates of $\hat{u}_{t+h}$ are far less correlated than those of $x_{t+h}$, so
that a penalized regression on them selects a stable and interpretable subset.

The third step selects. The naming regression keeps the common factors as unpenalized controls and
applies the sparsity penalty only to the decorrelated indicators,
\begin{equation}
  \xi^{\mathrm{b}}_{t+h}=\hat{f}_{t+h}^{\top}\alpha+\hat{u}_{t+h}^{\top}\theta+\nu_{t+h},
\label{eq:naming_reg}
\end{equation}
with common-factor loadings $\alpha\in\mathbb{R}^{\hat{r}}$ and idiosyncratic loadings
$\theta\in\mathbb{R}^{p}$, and the estimates solve
\begin{equation}
  (\hat{\alpha},\hat{\theta})
  =\argmin_{\alpha\in\mathbb{R}^{\hat{r}},\,\theta\in\mathbb{R}^{p}}\;
  \frac{1}{n}\sum_{t}\rho_\delta\!\Big(\xi^{\mathrm{b}}_{t+h}-\hat{f}_{t+h}^{\top}\alpha
      -\hat{u}_{t+h}^{\top}\theta\Big)
  +\sum_{j=1}^{p}p_\lambda(|\theta_j|),
\label{eq:farm_obj}
\end{equation}
the sum running over the $n$ naming origins. The penalty $p_\lambda$ is the SCAD penalty of
\citet{fan2001variable}, defined for $\theta\ge0$ by
\[
  p_\lambda(\theta)=
  \begin{cases}
    \lambda\theta, & 0\le\theta\le\lambda,\\[4pt]
    \dfrac{2a\lambda\theta-\theta^{2}-\lambda^{2}}{2(a-1)}, & \lambda<\theta\le a\lambda,\\[8pt]
    \dfrac{(a+1)\lambda^{2}}{2}, & \theta>a\lambda,
  \end{cases}
\]
with regularization strength $\lambda>0$ and shape parameter $a=3.7$.

SCAD is chosen because the object of interest is the identity and magnitude of the drivers. Being flat
beyond $a\lambda$, it estimates a driver that clears the threshold with asymptotically no shrinkage, so
that its loading $\hat{\theta}_j$ measures how strongly indicator $j$ moves the factor, a magnitude the
lasso would understate. It is also selection-consistent under the correlated design that
\eqref{eq:decorrelated} attenuates but does not remove, for which the lasso would require an
irrepresentable condition, so the selected name is stable across resamples.

The loss $\rho_\delta$ is the Huber loss \citep{huber1964robust}, and it enters for the same reason the
robust decision of Section~\ref{sec:framework} guards against headline-driven episodes: the target
$\xi^{\mathrm{b}}_{t+h}$ is a projected residual that, by construction, spikes on exactly such months.
The Huber loss,
\[
  \rho_\delta(c)=
  \begin{cases}
    c^{2}/2, & |c|\le\delta,\\[2pt]
    \delta|c|-\delta^{2}/2, & |c|>\delta,
  \end{cases}
\]
is quadratic for small residuals and linear beyond the threshold $\delta$, so that each month contributes
a gradient of magnitude at most $\delta$ and no single date can drive the selection while the estimator
stays efficient on the bulk of the sample. The threshold is set to the standard $\delta=1.345\,\hat{s}$,
with $\hat{s}$ the normalized median absolute deviation of the current residuals, updated at every
step of the fit, and the regularization strength
$\lambda$ is chosen over a fixed grid by ten-fold cross-validation, the factor extraction repeated
within every training fold and the held-out observations projected through the training-fold
loadings, so the selection criterion is itself out of sample.

The two coefficient blocks carry two kinds of economic content. A nonzero common-factor loading
$\hat{\alpha}_j$ ties the factor to the $j$th principal component of the panel, a diffuse economy-wide
direction rather than any single series, while a nonzero idiosyncratic coefficient $\hat{\theta}_j$ ties
it to indicator $j$ specifically, over and above the common factors, and so supplies a concrete name.
The pair $[\hat{\alpha}\mid\hat{\theta}]$ thus classifies the recovered factor as broad common
variation, a named individual driver, or a mixture of the two. Estimating \eqref{eq:farm_obj} separately
for the linear benchmark ($\mathrm{b}=\mathrm{L}$) and the forest ($\mathrm{b}=\mathrm{RF}$) names the
factor each leaves in its residual, and Section~\ref{subsec:emp_recovery} reads the difference between
the two selections as the structure the linear model leaves in its residual but the forest absorbs into
its forecast.

Both the target and the regressors in \eqref{eq:farm_obj} are estimated from the data. The target
$\xi^{\mathrm{b}}_{t+h}$ is produced by the forcing \eqref{eq:forcing} and the projection
\eqref{eq:projected_factor}; the regressors $\hat{f}_{t+h},\hat{u}_{t+h}$ are extracted from the
panel; the penalty is selected on the same sample. Classical $p$-values therefore do not apply to
the coefficients, and we assess significance by a block-permutation test.

The statistic $\hat{T}$ is the cross-validated out-of-sample $R^{2}$ at the selected penalty, the
criterion the naming itself maximizes. The null hypothesis is that the target and the panel are not
associated. We impose it by blocks. The $n$ naming origins are grouped into consecutive blocks of
$\ell\approx n^{1/3}$ months, and a permutation reorders the blocks of the target alone, holding
the panel fixed. The reordering destroys the alignment between target and indicators and preserves
the serial dependence of $\xi^{\mathrm{b}}_{t+h}$ within blocks.

The test runs as follows. We compute $\hat{T}$ on the original data and denote it
$\hat{T}^{\mathrm{obs}}$. If the cross-validated $R^{2}$ is not positive, the factor predicts no
better than a constant, and we declare it non-significant without drawing a permutation. Otherwise
we draw $L$ block permutations under a fixed seed, so the reported value is reproducible. On the
$q$th permuted target we re-run the penalty selection and the naming fit in full and record
$\hat{T}^{(q)}$. Because the selection is repeated on every permutation, the null distribution is
that of the best association attainable by chance under the same selection. The $p$-value is the
Monte Carlo estimate
\[
  \hat{p}=\frac{1+\#\{q\le L:\hat{T}^{(q)}\ge\hat{T}^{\mathrm{obs}}\}}{L+1},
\]
and small $\hat{p}$ rejects the null of no association.

\section{Empirical Study}
\label{sec:empirical}

We take the framework of Section~\ref{sec:framework} to U.S.\ Treasury benchmark yields. The forecasters of
Section~\ref{sec:benchmarks} supply the residuals, the forcing statistic \eqref{eq:forcing} recovers a
direction from those residuals, and the FarmSelect procedure of Section~\ref{sec:naming} names it. The
object of the study is the recovery and naming of the omitted factor, in
Section~\ref{subsec:emp_recovery}. The
forecasting results of Section~\ref{subsec:emp_forecast} enter only to establish that the residuals are
a sound input, and the decision baselines of Section~\ref{subsec:emp_baselines} locate the two-layer
decision against simpler rules for choosing the same exposure.

\subsection{Data}
\label{sec:data}
We use monthly, end-of-month zero-coupon U.S.\ Treasury benchmark yields from the LSEG Reuters
Workspace. The sample runs from January~2006 to August~2025, giving $T=236$ months, and covers
$N=15$ maturities from three months to thirty years. The period spans the global financial crisis,
the European sovereign debt crisis, the 2015--2018 normalization, the COVID-19 shock, and the
tightening cycle beginning in mid-2022, so the curve is far from stationary. The design
accommodates this instability directly: the forecasters of Section~\ref{sec:benchmarks} are
re-estimated every month on rolling windows of $w=60$ months, and the recovery of
Section~\ref{sec:framework} is built on a training set that expands with every month, using at
each origin every realized residual of its own forecaster. These two windows set the timeline of
the study. The rolling forecasts produce one-step residuals for target months from January~2016
onward, and the accuracy comparison of Section~\ref{subsec:emp_forecast} runs on the common window
of the two forecasters, the $115$ months from February~2016 through August~2025. The expanding
recovery deploys after its burn-in of $36$ residuals, January~2016 through December~2018, so the
deployed and naming months are the $80$ from January~2019 through August~2025, common to both
forecasters.

The predictor panel $Z_t\in\mathbb{R}^{p}$ consists of $p=111$ monthly indicators in three groups:
macroeconomic series spanning prices, labor, real activity, business surveys, and the household,
housing, and external sectors; Treasury supply-and-demand variables covering federal debt and securities
outstanding and Treasury capital flows; and financial variables comprising short-term and policy
interest rates together with the monetary aggregates. The full list is given in
Table~\ref{tab:EC_us_indicators} in Appendix~\ref{app:tables}. The factor-augmented forecaster of
Section~\ref{sec:benchmarks} takes each series after a stationarity transformation applied within every
estimation window: an augmented Dickey--Fuller test at the $10\%$ level differences the series once when
the unit-root null is not rejected and leaves it in levels otherwise \citep{dickey1979distribution}, after which the transformed series
is standardized to zero mean and unit variance on the window. The forest scales the raw predictors to
the unit interval within its window, as Section~\ref{sec:benchmarks} describes, and the naming
regression of Section~\ref{sec:naming} standardizes the panel over the naming sample. Every predictor
enters lagged by at least one month, matching its release calendar, so no forecast uses
contemporaneously unavailable information.

\subsection{Benchmark forecast performance}
\label{subsec:emp_forecast}

The recovery of the previous section reads the second moment of the forecast residuals, and the value of
what it recovers rests on those residuals being a genuine object rather than the artifact of a poorly
specified forecast. We therefore open the empirical study by locating the two forecasters of
Section~\ref{sec:benchmarks} against the no-change random walk
$\hat{y}^{\mathrm{RW}}_{t+1\mid t}(\tau)=y_t(\tau)$, which sets the forecast at each
maturity $\tau$ equal to the current yield and is the standard yardstick of this
literature \citep{duffee2002term}. Each forecaster is estimated in ten specifications, the random forest under ten
random seeds and the factor-augmented dynamic Nelson--Siegel model under the factor counts $k=1,\dots,10$,
and we carry all ten through the study rather than commit to one; both are evaluated over the same $115$
months.

Equality of one-step accuracy is tested by the statistic of \citet{diebold1995comparing}. Let
$e_{t+1}(\tau)=y_{t+1}(\tau)-\hat{y}_{t+1\mid t}(\tau)$ be a forecaster's error at maturity $\tau$ and
$d_t(\tau)=e_{t+1}(\tau)^2-e^{\mathrm{RW}}_{t+1}(\tau)^2$ its squared-error loss relative to the random
walk, with sample mean $\bar{d}(\tau)$, so that the null of equal expected accuracy is
$\mathbb{E}[d_t(\tau)]=0$. Under this null the statistic
\[
  \mathrm{DM}(\tau)=\frac{\bar{d}(\tau)}{\sqrt{\hat{\omega}(\tau)/n}}
\]
is asymptotically standard normal, where $n$ is the number of evaluation months and $\hat{\omega}(\tau)$
the Newey--West long-run variance of $d_t(\tau)$ truncated at lag $h-1$, which at the one-month horizon
is the sample variance of $d_t(\tau)$; we apply the small-sample correction of \citet{harvey1997testing}
and refer $\mathrm{DM}(\tau)$ to the $t_{n-1}$ distribution. A negative value indicates a forecaster more
accurate than the random walk and a positive value one less accurate, and
Table~\ref{tab:forecast_summary} reports the median and range of $\mathrm{DM}(\tau)$ across each
forecaster's ten specifications.

The two forecasters lie on opposite sides of the random walk. The random forest is more accurate at
every maturity of two years and beyond, where its median statistic is negative and, from four years out,
significant for all ten seeds; at maturities of one year and below the seeds straddle zero and the difference is not significant. The factor-augmented Nelson--Siegel model runs the other way. Its statistic is positive at
three months and at every maturity of two years and beyond, significant across all ten factor counts, so
the random walk is more accurate over most of the curve; only at six months and one year are the two
indistinguishable.

That the linear model does not beat the random walk is no defect here. Its forecast is linear in the
level, slope, and curvature of the curve and in the leading components of the indicator panel, and
cannot represent the nonlinear structure the recovery extracts. A linear model that beat the random walk
would be one whose span had already absorbed that structure, leaving less in the residual to recover.

\begin{table}[h]
\centering
\begin{threeparttable}
\caption{One-step ($h=1$) Diebold--Mariano tests against the no-change random walk, over $115$ months
(2016--2025).}
\label{tab:forecast_summary}
\begin{tabular}{lcc}
\toprule
$\tau$ & Random forest & FADNS \\
\midrule
3M  & $1.26\;[-1.04,2.00]$          & $2.27^{**}\;[2.02,2.61]$ \\
6M  & $0.93\;[-0.26,1.87]$          & $0.62\;[0.37,1.12]$ \\
1Y  & $0.20\;[-4.14,0.83]$          & $0.43\;[0.10,0.55]$ \\
2Y  & $-3.20\;[-4.16,-0.86]$        & $2.96^{***}\;[2.83,3.44]$ \\
3Y  & $-3.05\;[-5.03,-1.65]$        & $3.50^{***}\;[3.44,3.79]$ \\
4Y  & $-4.45^{***}\;[-5.56,-3.61]$  & $3.39^{***}\;[3.35,3.57]$ \\
5Y  & $-4.18^{**}\;[-5.90,-2.26]$   & $2.95^{***}\;[2.87,3.05]$ \\
6Y  & $-4.13^{**}\;[-5.85,-2.37]$   & $2.58^{**}\;[2.37,2.72]$ \\
7Y  & $-4.89^{***}\;[-5.85,-3.39]$  & $2.43^{**}\;[2.08,2.64]$ \\
8Y  & $-5.30^{***}\;[-6.18,-3.87]$  & $2.47^{**}\;[2.06,2.73]$ \\
9Y  & $-4.88^{**}\;[-5.92,-2.57]$   & $2.69^{**}\;[2.26,2.99]$ \\
10Y & $-4.88^{***}\;[-5.50,-3.23]$  & $3.27^{***}\;[2.79,3.52]$ \\
15Y & $-4.91^{***}\;[-6.16,-3.66]$  & $4.39^{***}\;[3.92,4.64]$ \\
20Y & $-4.86^{***}\;[-5.98,-4.17]$  & $4.40^{***}\;[3.96,4.62]$ \\
30Y & $-4.36^{***}\;[-6.19,-3.58]$  & $4.24^{***}\;[3.44,4.48]$ \\
\bottomrule
\end{tabular}
\begin{tablenotes}[flushleft]\footnotesize
\item Each cell gives the median and, in brackets, the $[\min,\max]$ range of the Diebold--Mariano
statistic $\mathrm{DM}(\tau)$ of Section~\ref{subsec:emp_forecast} across a forecaster's ten
specifications: ten random-seed refits for the random forest, factor counts $k=1,\dots,10$ for FADNS. A
negative value indicates the forecaster is more accurate than the random walk, a positive value less
accurate. Stars mark rejection of equal accuracy for all ten specifications at the stated level
($^{*}p<0.10$, $^{**}p<0.05$, $^{***}p<0.01$). Per-specification statistics are in
Table~\ref{tab:dm_byspec} in Appendix~\ref{app:tables}.
\end{tablenotes}
\end{threeparttable}
\end{table}

\subsection{The recovery as a Sharpe decision}
\label{subsec:emp_baselines}

Each decision rule in this subsection produces a duration position $\omega\in\mathbb{R}^{N}$ across the
maturities, scaled to unit ex-ante volatility and scored by the out-of-sample Sharpe ratio of its
profit and loss
\begin{equation}
  \Pi^{\mathrm{b}}_{t+h}(\omega)=-\,\omega^{\top}\Delta y_{t+h},
  \qquad \Delta y_{t+h}=y_{t+h}-y_t .
\label{eq:pnl}
\end{equation}
All rules share the residual covariance $\hat{\Sigma}^{\mathrm{b}}_{z,t}$ as risk model and the
forecast-implied mean input
\begin{equation}
  \hat{\mu}^{\mathrm{b}}_t=-\,\widehat{\Delta y}^{\mathrm{b}}_{t+h}
  =-\big(\hat{y}^{\mathrm{b}}_{t+h\mid t}-y_t\big),
\label{eq:mean_input}
\end{equation}
the negative of the benchmark's forecast yield change, so a position profits in expectation where the
forecast calls yields lower; the rules differ only in that mean input or in the two radii. The
single-model rule is the nominal maximum-Sharpe exposure on the benchmark's own mean input
($\gamma_1=0$, $\rho=0$), the baseline exposure $\omega^{\mathrm{b},\ast}_t(0)$
from which Section~\ref{subsec:emp_recovery} recovers the factor. The combination rule replaces that
mean input by a pool-combined forecast under each of the eleven schemes of
Appendix~\ref{app:combination}, the ten random-forest seeds combined among themselves and the ten FADNS
factor counts among themselves. The one- and two-layer rules restore the mean layer of
\eqref{eq:delage_ye_set} and the mean and covariance layers together.

The ambiguity set \eqref{eq:delage_ye_set} has three parameters: the mean-ellipsoid radius $\gamma_1$, the
covariance-scaling factor $\gamma_2$, and the covariance-inflation radius $\rho$. We fix $\gamma_2=1$ and
calibrate $\gamma_1$ and $\rho$ from the data of each training set, following
\citet{delage2010distributionally}, so that the robustness the two layers impose is set by the sample
rather than chosen.

Both radii are computed on the expanding training set of length $M_t$ ending at the forecast origin $t$,
beginning after a burn-in of $36$ months, where the standardized shocks $z_s$ of \eqref{eq:std_residual}
and their shrunk covariance $\hat{\Sigma}^{\mathrm{b}}_{z,t}$ of \eqref{eq:shrunk_cov} are already in
hand. Two statistics of the training set enter
the radii. The first is the support radius $R_t$, the smallest radius of a ball (in the
$\hat{\Sigma}^{\mathrm{b}}_{z,t}$-Mahalanobis metric) containing the standardized shocks; to avoid
sensitivity to a single outlier we set $R^2_t$ to the $95$th percentile of the squared Mahalanobis norms
$\{z_s^{\top}(\hat{\Sigma}^{\mathrm{b}}_{z,t})^{-1}z_s\}_{s}$ over the training set. The second is the
average eigenvalue of the covariance,
$\bar{\lambda}_t=N^{-1}\operatorname{tr}\hat{\Sigma}^{\mathrm{b}}_{z,t}$, where $N=15$ is the number of
maturities.

\emph{Mean layer.} The mean-ellipsoid radius $\gamma_{1,t}$ is the finite-sample confidence bound of
\citet[Corollary~1]{delage2010distributionally}, under which the true shock mean lies in the ellipsoid
$\{m:m^{\top}(\hat{\Sigma}^{\mathrm{b}}_{z,t})^{-1}m\le\gamma_{1,t}\}$ with probability at least $1-\delta$:
\begin{equation}
  \gamma_{1,t}=\frac{R^2_t}{M_t}\Big(2+\sqrt{2\ln(1/\delta)}\Big)^2,\qquad \delta=0.05 .
\label{eq:gamma1_calib}
\end{equation}
It grows with the support radius $R_t$ and the confidence level $1-\delta$, and shrinks as $1/M$ with the
training length.

\emph{Covariance layer.} The inflation radius $\rho_t$ guards the dominance bound of
the ambiguity set against sampling error in $\hat{\Sigma}^{\mathrm{b}}_{z,t}$: finite
samples spread the sample eigenvalues around their population values, and it is the
downward-biased small ones that a bound built on $\hat{\Sigma}^{\mathrm{b}}_{z,t}$
alone would inherit. We therefore lift the whole spectrum by the total standardized
variance per training observation,
\begin{equation}
  \rho_t=\frac{N}{M_t}\,\bar{\lambda}_t=\frac{\operatorname{tr}\hat{\Sigma}^{\mathrm{b}}_{z,t}}{M_t},
\label{eq:rho_calib}
\end{equation}
which likewise shrinks as $1/M$. Both radii are recomputed every month.

\begin{table}[h]
\centering
\begin{threeparttable}
\caption{Out-of-sample Sharpe ratio of the duration decision: single model, forecast combination,
and one- and two-layer robustness.}
\label{tab:sharpe_decision}
\begin{tabular}{lcc}
\toprule
Rule & Random forest & FADNS \\
\midrule
Single model                  & $1.153$ & $0.533\;[0.499,0.554]$ \\
DRO, one layer                & $0.952$ & $0.527\;[0.498,0.539]$ \\
DRO, two layers               & $0.967$ & $0.530\;[0.497,0.542]$ \\
\midrule
Combination, EW               & $1.183$ & $0.545$ \\
Combination, RANK             & $1.192$ & $0.555$ \\
Combination, RMSE             & $1.185$ & $0.545$ \\
Combination, MSE              & $1.215$ & $0.572$ \\
Combination, OLS              & $1.189$ & $0.547$ \\
Combination, MV               & $1.234$ & $0.548$ \\
Combination, STACK            & $1.156$ & $0.568$ \\
Combination, LAD              & $1.170$ & $0.547$ \\
Combination, AFTER-Rolling    & $1.128$ & $0.535$ \\
Combination, AFTER-EWMA       & $1.166$ & $0.525$ \\
Combination, AFTER-Simplified & $1.205$ & $0.544$ \\
\bottomrule
\end{tabular}
\begin{tablenotes}[flushleft]\footnotesize
\item Realized Sharpe ratio of $\Pi^{\mathrm{b}}_{t+h}=-\omega^{\top}\Delta y_{t+h}$, the exposure scaled
to unit ex-ante volatility each month, on the expanding scheme with a $36$-month burn-in. The FADNS column
gives the median and $[\min,\max]$ over $k=1,\dots,10$ for the single-model and robust rules, and a single
value per combination scheme, whose pool combines the ten counts into one series. The random-forest
column uses the seed-averaged forecast for the single-model and robust rules, and the ten seeds pooled
into one combined series for the combination rules. Single-model and
combination rules are nominal ($\gamma_1=0$, $\rho=0$); the robust rules use $\gamma_{1,t}$ and $\rho_t$ of
\eqref{eq:gamma1_calib}--\eqref{eq:rho_calib}.
\end{tablenotes}
\end{threeparttable}
\end{table}

Table~\ref{tab:sharpe_decision} shows three things. First, the nominal exposure is already a strong Sharpe
decision, the forest at $1.15$ and FADNS at a median of $0.53$. Second, the robust layers do not improve
on it: they lower the forest to $0.95$ and $0.97$ and leave FADNS essentially unchanged at $0.53$, the
second layer moving the ratio only slightly in either case. Third, combination improves both forecasters,
but mildly---the forest to between $1.13$ and $1.23$ and FADNS to between $0.53$ and $0.57$ across the
eleven schemes---so no rule moves the decision far from the nominal exposure.

That the robust layers cost rather than add Sharpe is expected. At $\rho=0$ the reduced program
\eqref{eq:robust_sharpe_reduced} has optimal value
$\lVert (D^{\mathrm{b}}_t)^{-1}\hat{\mu}^{\mathrm{b}}_t\rVert_{(\hat{\Sigma}^{\mathrm{b}}_{z,t})^{-1}}-\sqrt{\gamma_{1,t}}$,
the attainable Sharpe net of the mean penalty, so the exposure is withdrawn whenever the standardized
signal falls short of $\sqrt{\gamma_{1,t}}$, trading realized return for protection the Sharpe does not
reward; the covariance layer, calibrated to estimation error rather than to a stress, tilts the exposure
only slightly. The mean--variance criterion measured here is therefore not where the framework earns its
keep: the two-layer exposure returns in Section~\ref{sec:dv01} as the deployed book, and the
baseline exposure $\omega^{\mathrm{b},\ast}_t(0)$ is the input from which the next subsection recovers
the factor.

\subsection{The recovered factor and its name}
\label{subsec:emp_recovery}

The recovery runs on the same expanding scheme as the decision study. At each origin $t$ the direction
$v^{\mathrm{b}}_t$ of \eqref{eq:forcing} is estimated on all months up to $t$, beginning after the
$36$-month burn-in, and the factor is the out-of-sample projection
$\xi^{\mathrm{b}}_{t+1}=v^{\mathrm{b}\top}_t z_{t+1}$, where the shock $z_{t+1}$ is standardized with
training moments only, so that no quantity entering $\xi^{\mathrm{b}}_{t+1}$ uses information beyond $t$.
The forest enters through its seed-averaged forecast, one residual object, while each FADNS factor count
is its own, so the recovery and the naming run once for the forest and once per count. This yields $80$
out-of-sample observations per benchmark over 2019--2025. The naming regression of
Section~\ref{sec:naming} then relates $\xi^{\mathrm{b}}$ to the $111$-indicator panel, with the penalty
chosen by cross-validation and significance assessed by the block-permutation test with nested penalty
selection on $199$ draws; the eigenvalue-ratio estimate selects a single common factor, $\hat{r}=1$, in
every specification. Since the regression relates $n=80$ observations to $p=111$ indicators, it is a
regularized association in the $p>n$ regime, and we read its selections as names rather than as a
structural model.

\begin{table}[h]
\centering
\begin{threeparttable}
\caption{Naming the recovered factor: leading indicators, fit, and permutation significance.}
\label{tab:naming}
\footnotesize
\begin{tabular}{lp{0.47\textwidth}ccc}
\toprule
Spec. & Leading selected indicators & OOS $R^2$ & Sel. & $p$ \\
\midrule
RF     & private inventories; M2; building permits; Chicago PMI & $0.210$ & $4$  & $0.135$ \\
\midrule
$k=1$  & \textbf{leading index}; dollar--euro; dollar index; ext.\ balance & $0.242$ & $4$  & $0.020^{**}$ \\
$k=2$  & private inventories & $0.290$ & $1$  & $0.010^{***}$ \\
$k=3$  & \textbf{leading index}; dollar--euro; private inventories & $0.315$ & $3$  & $0.005^{***}$ \\
$k=4$  & \textbf{leading index}; funds rate; Treasury debt; M2; \dots & $0.265$ & $18$ & $0.020^{**}$ \\
$k=5$  & \textbf{leading index}; funds rate; Treasury debt; M2; \dots & $0.265$ & $19$ & $0.020^{**}$ \\
$k=6$  & \textbf{leading index}; funds rate; job cuts; permits; \dots & $0.124$ & $9$  & $0.075^{*}$ \\
$k=7$  & \textbf{leading index}; funds rate; job cuts; foreign flows; \dots & $0.143$ & $8$  & $0.050^{**}$ \\
$k=8$  & \textbf{leading index}; funds rate; job cuts; foreign flows; \dots & $0.152$ & $7$  & $0.050^{**}$ \\
$k=9$  & \textbf{leading index}; interbank rate; confidence; job cuts; \dots & $0.168$ & $10$ & $0.040^{**}$ \\
$k=10$ & \textbf{leading index}; interbank rate; confidence; job cuts; \dots & $0.170$ & $10$ & $0.065^{*}$ \\
\bottomrule
\end{tabular}
\begin{tablenotes}[flushleft]\footnotesize
\item For each specification: the selected indicators with the largest loadings (the Conference Board
leading index in bold where selected; ``\dots'' marks further selections listed in full, with
coefficients, in Table~\ref{tab:naming_full}), the out-of-sample $R^2$ of the naming regression, the
number of indicators selected, and the exact permutation $p$-value on $199$ draws with nested penalty
selection ($^{*}p\le0.10$, $^{**}p\le0.05$, $^{***}p\le0.01$). All specifications use the $80$
out-of-sample factor observations, and the penalty is chosen by cross-validation per specification.
\end{tablenotes}
\end{threeparttable}
\end{table}

Table~\ref{tab:naming} reports the fit and the permutation evidence. For the linear benchmark the factor
names: eight of the ten factor counts reject the permutation null at the five-percent level and all ten at
ten percent, with out-of-sample $R^2$ between $0.12$ and $0.32$, and although the number of selected
indicators varies from one at $k=2$ to nineteen at $k=5$, the selections share a stable core. The
Conference Board leading index is selected in nine of the ten specifications and carries the largest or
near-largest loading in each; a policy-rate variable accompanies it from $k=4$ onward, the federal funds
target rate through $k=8$ and the three-month interbank rate thereafter; and a small set of real-activity
and external-account indicators---job-cut announcements, building permits, foreign net long-term
securities flows, the goods trade balance---recurs across the larger selections, whose full composition
Table~\ref{tab:naming_full} in Appendix~\ref{app:tables} lists. Because the overall sign of each specification's factor is arbitrary,
the leading index entering with opposite signs at different $k$ carries no information; what is
informative is its recurrence. Taken together, the evidence names the factor in the linear residual as a
leading business-cycle factor, anchored on the leading index and accompanied by the policy stance and the
external accounts.

The forest behaves differently, and in the direction the forecast comparison of
Section~\ref{subsec:emp_forecast} would predict. Its naming regression attains an out-of-sample $R^2$ of
$0.21$ and selects four indicators---the M2 money stock, the Chicago purchasing managers' index, building
permits, and the change in private inventories---each of which the linear specifications at $k=4$ and
$k=5$ also select, yet the permutation test does not certify the association, returning $p=0.135$. A
forest that beats the random walk has absorbed more of the systematic content into its forecast, and the
trace it leaves in the residual, while pointing at the same real-activity family, is too faint to name at
this sample length.

The two recoveries are accordingly related but not identical. The angle between the averaged directions
$v^{\mathrm{L}}$ and $v^{\mathrm{RF}}$ rises from $39$ degrees at $k=1$ to $73$ at $k=5$ and settles near
$66$ thereafter, and the selected indicator sets overlap in at most four names
(Table~\ref{tab:contrast} in Appendix~\ref{app:tables}); the two benchmarks leave different second-moment geometry in their residuals,
and the recovery reads each on its own terms.

\section{Decision Application: DRO for DV01 Risk Management}
\label{sec:dv01}

The framework of Section~\ref{sec:framework} is stated for a generic decision, and the study of
Section~\ref{subsec:emp_baselines} already deployed it as a duration position with profit and loss
\eqref{eq:pnl} under the forecast-implied mean input \eqref{eq:mean_input}. This section reads that
position as interest-rate risk management. Read one way the robust duration position is the decision a
risk manager deploys, whose out-of-sample value as a Sharpe decision Section~\ref{subsec:emp_baselines}
established; read the other way its closed-form member is the baseline exposure that indexes the
covariance forcing \eqref{eq:forcing} and thereby recovers the omitted factor. This section develops the
second reading into a risk-management operation: neutralizing the recovered factor from the deployed
book, and asking what that neutralization does to the tail of the realized profit and loss. The answer
is itself evidence about the factor.

\subsection{The duration book and its residual risk}
\label{subsec:dv01_decision}

A risk manager holds a book of Treasury exposures across the $N=15$ maturities. Let
$\omega\in\mathbb{R}^{N}$ collect the position's key-rate DV01, so that $\omega_\tau$ is the profit and
loss per one-basis-point move in the maturity-$\tau$ yield; long and short exposures are allowed, so
$\omega$ is unconstrained in sign. Over the horizon $h$ the realized and forecast yield changes are
\begin{equation}
  \Delta y_{t+h} = y_{t+h}-y_t\in\mathbb{R}^{N},
  \qquad
  \widehat{\Delta y}^{\mathrm{b}}_{t+h} = \hat{y}^{\mathrm{b}}_{t+h\mid t}-y_t,
  \qquad
  \Delta y_{t+h} = \widehat{\Delta y}^{\mathrm{b}}_{t+h}+e^{\mathrm{b}}_{t+h},
\label{eq:yield_change_decomp}
\end{equation}
so the realized change is the benchmark forecast of the change plus the residual $e^{\mathrm{b}}_{t+h}$
of Section~\ref{subsec:setup}. To first order the profit and loss \eqref{eq:pnl} accordingly splits as
\[
  \Pi^{\mathrm{b}}_{t+h}(\omega)
  = \underbrace{-\,\omega^{\top}\widehat{\Delta y}^{\mathrm{b}}_{t+h}}_{\text{forecast-implied mean}}
    \;\underbrace{-\;\omega^{\top} e^{\mathrm{b}}_{t+h}}_{\text{residual risk}} ,
\]
and the split supplies exactly the two inputs the robust program requires. The first term is
$\omega^{\top}\hat{\mu}^{\mathrm{b}}_t$ under the mean input \eqref{eq:mean_input}, and the risk is
carried by the residual, whose covariance in raw units is
$\Sigma^{\mathrm{b}}_{e,t}=D^{\mathrm{b}}_t\,\hat{\Sigma}^{\mathrm{b}}_{z,t}\,D^{\mathrm{b}}_t$ with
$\hat{\Sigma}^{\mathrm{b}}_{z,t}$ and $D^{\mathrm{b}}_t$ as in
\eqref{eq:std_residual}--\eqref{eq:shrunk_cov}. With the volatility-scaled exposure
$a^{\mathrm{b}}_t=D^{\mathrm{b}}_t\,\omega$ of \eqref{eq:robust_sharpe}, the residual term satisfies
$(a^{\mathrm{b}}_t)^{\top}z^{\mathrm{b}}_{t+h}=\omega^{\top}\big(e^{\mathrm{b}}_{t+h}-\bar{e}^{\mathrm{b}}_t\big)$
and
$\operatorname{Var}(\omega^{\top}e^{\mathrm{b}}_{t+h})=(a^{\mathrm{b}}_t)^{\top}\hat{\Sigma}^{\mathrm{b}}_{z,t}\,a^{\mathrm{b}}_t$,
so that $\Pi^{\mathrm{b}}_{t+h}(\omega)$ has the form of the framework's position return
\eqref{eq:position_return} under the mean input \eqref{eq:mean_input}, and the covariance forcing
\eqref{eq:forcing} is computed in the standardized residual space of a genuine duration decision. The
shock $z^{\mathrm{b}}$ enters the ambiguity set only through a symmetric mean ellipsoid and a covariance
bound, so its orientation is immaterial to both the robust program and the forcing. Because the forecast
enters only through the mean input \eqref{eq:mean_input}, holding the forecasting model fixed is exactly
the post-prediction premise of the framework: the manager keeps the forecast and robustifies the residual
risk the position carries.

\subsection{The robust duration position}
\label{subsec:robust_position}

Substituting the mean input \eqref{eq:mean_input} into the position return \eqref{eq:position_return}
and taking the two-layer moment set $\mathcal{A}^{\mathrm{b}}_t(\rho)$ of \eqref{eq:delage_ye_set}
places the duration decision in the worst-case Sharpe program \eqref{eq:robust_sharpe}. Its reduction
and its solution are those of Section~\ref{subsec:two_layer}: the reduced program
\eqref{eq:robust_sharpe_reduced} is solved in normalized form at each origin of the expanding scheme of
Section~\ref{subsec:emp_baselines}, in closed form at $\gamma_1=0$ and as the norm-constrained concave
program for $\gamma_1>0$, the branch that runs here. The objective is scale-invariant in $\omega$, and
the deployed exposure is scaled to unit ex-ante volatility at every origin.

Two exposures act in what follows, and they are distinct. The recovery baseline of
Sections~\ref{sec:framework} and~\ref{subsec:emp_recovery} is the closed-form case
$a^{\mathrm{b}}_t=D^{\mathrm{b}}_t\,\omega^{\mathrm{b},\ast}_t(0)$ at $\gamma_1=0$ and $\rho=0$. The
deployed book is the two-layer exposure of Section~\ref{subsec:emp_baselines}: the worst-case-Sharpe
direction under the calibrated radii $\gamma_{1,t}$ and $\rho_t$ of
\eqref{eq:gamma1_calib}--\eqref{eq:rho_calib}, scaled to unit ex-ante volatility at every origin. At an
origin where the robust numerator is nonpositive the position is withdrawn, as in
Section~\ref{subsec:two_layer}, and that month is excluded for all three positions alike, so the
profit-and-loss series compared below share identical dates. The forcing \eqref{eq:forcing} applies at
any fixed exposure, and the neutralization below evaluates it at the deployed book itself, so the
direction removed is the residual-risk direction this book carries but the forecast does not span.

\subsection{Neutralizing the recovered factor}
\label{subsec:dv01_hedge}

The recovered direction identifies a specific residual exposure carried by the deployed book, and the
position can be neutralized against it. The natural measure of that exposure is not the geometric overlap
of $v^{\mathrm{b}}_t$ with the exposure but the covariance between the position's residual return and the
realized factor. Because the forcing statistic is orthogonal to the exposure by construction, the raw
inner product $(v^{\mathrm{b}}_t)^{\top}a^{\mathrm{b}}_t$ vanishes identically and cannot serve as a
diagnostic; the exposure is instead borne through the second moment,
\begin{equation}
  \kappa^{\mathrm{b}}_t
  = \operatorname{Cov}\!\big(\xi^{\mathrm{b}}_{t+h},\,(a^{\mathrm{b}}_t)^{\top}z^{\mathrm{b}}_{t+h}\big)
  = (v^{\mathrm{b}}_t)^{\top}\hat{\Sigma}^{\mathrm{b}}_{z,t}\,a^{\mathrm{b}}_t
  = \big\lVert \hat{d}^{\mathrm{b}}_t\big\rVert_2 ,
\label{eq:recovered_loading}
\end{equation}
where $\xi^{\mathrm{b}}_{t+h}=(v^{\mathrm{b}}_t)^{\top}z^{\mathrm{b}}_{t+h}$ is the realized factor of
\eqref{eq:projected_factor}, $v^{\mathrm{b}}_t$ is the forcing direction \eqref{eq:forcing} evaluated at
the deployed exposure $a^{\mathrm{b}}_t$, and the final equality holds because
$\hat{d}^{\mathrm{b}}_t=P^{\perp}_{a^{\mathrm{b}}_t}\hat{\Sigma}^{\mathrm{b}}_{z,t}a^{\mathrm{b}}_t$
satisfies $(\hat{d}^{\mathrm{b}}_t)^{\top}a^{\mathrm{b}}_t=0$, so the covariance loading of the exposure on
the recovered factor equals the norm of the very forcing statistic that recovered it. A manager who wishes
to remove this exposure imposes on the book the covariance-neutrality constraint
\begin{equation}
  \big(D^{\mathrm{b}}_t\,\hat{\Sigma}^{\mathrm{b}}_{z,t}\,v^{\mathrm{b}}_t\big)^{\top}\omega = 0,
\label{eq:hedge_constraint}
\end{equation}
which sets $\kappa^{\mathrm{b}}_t=0$ and so makes the position's residual return uncorrelated with the
recovered factor while leaving the forecast mean input \eqref{eq:mean_input} unchanged. We compare the
deployed book with two hedged variants of it, both formed by projecting the exposure onto the
covariance-neutral hyperplane \eqref{eq:hedge_constraint}: one renormalizes the projected exposure to
the original unit volatility budget, and one leaves it at the reduced volatility the projection implies.
The second variant isolates the effect of removing the factor from the effect of the renormalization
that removal induces.

\subsection{The tail under neutralization}
\label{subsec:dv01_empirics}

We deploy the three positions on the expanding scheme of Section~\ref{subsec:emp_baselines}: the book
and its renormalized hedge are scaled to unit ex-ante volatility at every origin, and the fixed-budget
hedge is left at the volatility its projection implies. We record the realized volatility, the maximum
drawdown of cumulative profit and loss, the five-percent conditional value at risk \citep{RockafellarUryasev2000}, and the mean
covariance loading $\bar{\kappa}$ of \eqref{eq:recovered_loading}, which both hedged variants set to
zero by construction. Table~\ref{tab:dv01_hedge} reports the outcome.

\begin{table}[h]
\centering
\begin{threeparttable}
\caption{Tail of the deployed two-layer book before and after neutralizing the recovered factor.}
\label{tab:dv01_hedge}
\footnotesize\setlength{\tabcolsep}{3pt}
\begin{tabular}{lcccc cccc}
\toprule
& \multicolumn{4}{c}{Random forest} & \multicolumn{4}{c}{FADNS (median $[\min,\max]$ over $k$)} \\
\cmidrule(lr){2-5}\cmidrule(lr){6-9}
Position & Vol. & MaxDD & $\mathrm{CVaR}_{5\%}$ & $\bar{\kappa}$
         & Vol. & MaxDD & $\mathrm{CVaR}_{5\%}$ & $\bar{\kappa}$ \\
\midrule
Two-layer            & $3.07$ & $4.93$ & $1.41$ & $1.24$
                     & $1.45\;[1.42,1.47]$ & $2.56\;[1.89,3.00]$ & $1.13\;[0.83,1.49]$ & $1.02\;[0.94,1.06]$ \\
Hedged, renormalized & $3.01$ & $9.28$ & $2.34$ & $0$
                     & $1.38\;[1.33,1.46]$ & $3.07\;[1.77,3.63]$ & $1.25\;[1.12,1.41]$ & $0$ \\
Hedged, fixed budget & $2.98$ & $8.42$ & $2.13$ & $0$
                     & $1.34\;[1.23,1.46]$ & $2.56\;[1.49,3.28]$ & $1.17\;[1.06,1.30]$ & $0$ \\
\bottomrule
\end{tabular}
\begin{tablenotes}[flushleft]\footnotesize
\item The deployed book and the renormalized hedge are scaled to unit ex-ante volatility at every origin
of the expanding scheme ($36$-month burn-in, $80$ months deployed); the fixed-budget hedge applies the
projection without renormalization. Vol.\ is the realized standard deviation of monthly profit
and loss, MaxDD the maximum drawdown of its cumulative sum, and $\mathrm{CVaR}_{5\%}$ the negative mean
of the worst five percent of months; $\bar{\kappa}$ is the mean covariance loading
\eqref{eq:recovered_loading}, zero under the hedge by construction. The FADNS columns report the median
and $[\min,\max]$ over the factor counts $k=1,\dots,10$.
\end{tablenotes}
\end{threeparttable}
\end{table}

For the random-forest benchmark the hedge leaves the realized volatility essentially
unchanged, $2.98$ to $3.01$ against $3.07$, while both tail measures deteriorate substantially: the
maximum drawdown rises from $4.93$ to $9.28$ under renormalization and to $8.42$ at a fixed budget, and
the five-percent CVaR from $1.41$ to $2.34$ and $2.13$ respectively. Since the fixed-budget variant
removes the factor exposure without the rescaling that removal induces, and the deterioration persists
under it, the effect is attributable to the removal itself rather than to amplification. We conclude that
over this sample the position's return was stabilized in its worst months through its covariance with the
recovered factor, so that imposing zero covariance eliminates a component that was offsetting tail
losses.

The linear benchmark presents a different picture. Although its loading is comparable in size, with
$\bar{\kappa}$ near one against $1.24$ for the forest, the tail consequence is smaller and less uniform:
at a fixed budget the median maximum drawdown across the ten factor counts is unchanged at $2.56$ and the
median CVaR rises from $1.13$ to $1.17$, with the sign of the change varying across $k$, while under
renormalization the median drawdown rises from $2.56$ to $3.07$. This asymmetry is consistent with the
forecast comparison of Section~\ref{subsec:emp_forecast} and the naming results of
Section~\ref{subsec:emp_recovery}: the forest absorbs more of the systematic content into its forecast,
and the factor remaining in its residual, although too faint to name at this sample length, is the one
whose neutralization the realized tail penalizes most. In both cases the recovered direction behaves as
structure rather than estimation noise, and the covariance-neutrality constraint
\eqref{eq:hedge_constraint} is accordingly not a protective overlay but the removal of a systematic
exposure whose contribution to the book was, over this sample, stabilizing.

\section{Conclusion}
\label{sec:conclusion}

The paper recovered an omitted factor from the forecast residuals of the U.S.\ Treasury curve and
returned it to the decision the forecast serves. The forecaster is held fixed, a duration exposure
is made robust over a two-layer moment ambiguity set on the standardized residual cross-section,
and the recovered direction is the covariance forcing at that exposure. The direction is invariant
to shrinkage and to isotropic inflation, identifies a single omitted loading up to sign and an
angle bound, and decomposes over several factors as an exposure-weighted mixture
(Propositions~\ref{prop:shrinkage_invariance}--\ref{prop:multifactor}): a property of the residuals
rather than of the regularization, dependent on the forecaster only through its residual
covariance.

The empirical study read one finding from two sides. On the residuals of the linear
factor-augmented dynamic Nelson--Siegel benchmark, a forecaster the random walk outperforms over
most of the curve, the factor names as a leading business-cycle factor, led by the Conference Board
leading index and certified by the block-permutation test in nearly every specification; on the
residuals of the more accurate random forest, the same procedure selects the same real-activity
family without certification. The restrictive model leaves the structure in its residual, where the
recovery is sharpest; the flexible model absorbs much of it into its forecast. The robust layers do
not raise the realized Sharpe ratio; they define the deployed book and index the recovery.
Neutralizing the recovered factor on that book worsens the tail it had been offsetting: the
direction is systematic structure, not sampling noise (Section~\ref{sec:dv01}).

Two limitations bound the claims: with $N=15$ maturities the recovery rests on the finite-sample
guarantees of Section~\ref{sec:framework}, not on high-dimensional asymptotics; with $80$ months
against $111$ indicators the naming is a regularized association under a permutation null, not a
structural model. Two directions remain. In high-dimensional cross-sections, where the series count
is commensurate with the training length and the panel mixes data of different types, the
invariance makes the forcing a natural statistic, but the identification conditions need
high-dimensional counterparts. Replacing the moment set with a Wasserstein ball
\citep{MohajerinEsfahani2018,kuhn2019wasserstein} asks whether an invariance of the recovered
direction survives ambiguity that is metric rather than moment-based.

\appendix

\section{Estimation of the benchmark forecasters}
\label{app:benchmarks}

This appendix collects the estimation and forecasting detail condensed from
Section~\ref{sec:benchmarks}. All quantities are computed on the rolling window of $w=60$ months and
re-estimated as the window advances one month, and forecasts are out of sample at the one-month
horizon $h=1$. The window index set is $\mathcal{I}_t=\{t-59,\ldots,t\}$.

\subsection{DNS and FADNS}
\label{app:fadns}

At each $t$ the level, slope, and curvature factors $\beta_t=(\beta_{1t},\beta_{2t},\beta_{3t})^{\top}$
in \eqref{eq:dns} are fitted by cross-sectional least squares over the $N=15$ maturities, with the
decay fixed at $\lambda=0.0609$. Under plain DNS the factors follow a first-order vector
autoregression, $\beta_{t+1}=c+\Phi\beta_t+\eta_t$ with $\eta_t\sim(0,\Sigma_\eta)$, estimated on the
window, and the one-month forecast is $\hat{\beta}_{t+h\mid t}=c+\Phi\beta_t$.

FADNS augments the state with the lagged panel $Z_t\in\mathbb{R}^{p}$, $p=111$, of macroeconomic,
Treasury supply-and-demand, and financial indicators described in Section~\ref{sec:benchmarks} and
listed in Table~\ref{tab:EC_us_indicators}. Within each window every predictor is tested for a unit
root by the augmented Dickey--Fuller test. A series is differenced once when the null is not rejected
at the $10\%$ level and kept in levels otherwise, and the transformed series is then standardized to
zero mean and unit variance, giving $\tilde{Z}_{t-j}$, $j=1,\ldots,w$. Only lagged predictors enter,
reflecting a one-month release lag. Principal component analysis is applied to the standardized block
$\{\tilde{Z}_{t-w},\ldots,\tilde{Z}_{t-1}\}$, with the eigenpairs of the block covariance ordered by
decreasing eigenvalue and their signs aligned to the previous window. The $j$th component available at
$t$ uses only information dated $t-1$ and earlier, $\mathrm{PC}_{j,t}=v_{j,t}^{\top}\tilde{Z}_{t-1}$,
and the panel-factor vector is $F^{(k)}_t=(\mathrm{PC}_{1,t},\ldots,\mathrm{PC}_{k,t})^{\top}$. The
augmented state $X^{(k)}_t=(\beta_{1t},\beta_{2t},\beta_{3t},F^{(k)\top}_t)^{\top}\in\mathbb{R}^{3+k}$
follows a first-order vector autoregression
$X^{(k)}_{t+1}=c^{(k)}+\Phi^{(k)}X^{(k)}_t+\eta^{(k)}_t$, estimated on the window. The one-month state
forecast is $\hat{X}^{(k)}_{t+h\mid t}=c^{(k)}+\Phi^{(k)}X^{(k)}_t$, whose first three coordinates give
$\hat{\beta}_{t+h\mid t}$, which \eqref{eq:dns} maps to $\hat{y}^{\mathrm{L}}_{t+h\mid t}(\tau)$ and the
residual \eqref{eq:resid_fadns}. No single factor count is selected. Each $k\in\{1,\ldots,10\}$ defines
its own specification, with the state $X^{(k)}$ estimated and the forecast formed on every window
exactly as above, and the empirical study of Section~\ref{sec:empirical} carries all ten as a range
rather than committing to one; $k=0$ is the plain-DNS case.

\subsection{Random forest}
\label{app:rf}

The response is the yield vector $y_{t+h}\in\mathbb{R}^{N}$ and the predictor is the scaled vector
$W_t\in[0,1]^{d_W}$ of \eqref{eq:rf_predictor} and the min--max map of Section~\ref{subsec:rf_family};
the response enters in its original units, so no inverse scaling is needed. The forest estimates the
multi-output regression \eqref{eq:rf} by averaging $B$ regression trees as in \eqref{eq:rf_ensemble},
each grown by the classification-and-regression-tree (CART) rule of recursive axis-aligned
partitioning.

A single tree partitions the unit cube $[0,1]^{d_W}$ into disjoint axis-aligned cells, its terminal
nodes or leaves, splitting one coordinate at a time. For a node $A\subseteq[0,1]^{d_W}$ with training
indices $\mathcal{I}_A=\{s\in\mathcal{I}_t:W_s\in A\}$, the node prediction is the vector mean
$\bar{y}_A=|\mathcal{I}_A|^{-1}\sum_{s\in\mathcal{I}_A} y_{s+h}\in\mathbb{R}^{N}$, and the node impurity
is the total within-node squared error across maturities,
\begin{equation}
  \mathrm{Imp}(A)=\frac{1}{|\mathcal{I}_A|}\sum_{s\in\mathcal{I}_A}\lVert y_{s+h}-\bar{y}_A\rVert_2^{2}.
\label{eq:rf_impurity}
\end{equation}
A split of $A$ on coordinate $j\in\{1,\ldots,d_W\}$ at threshold $c\in(0,1)$ forms the children
$A_L=\{W\in A:(W)_j\le c\}$ and $A_R=A\setminus A_L$, sending training index $s$ to $A_L$ when
$(W_s)_j\le c$, and is chosen to maximize the impurity reduction
\[
  \mathrm{Imp}(A)-\frac{|\mathcal{I}_{A_L}|}{|\mathcal{I}_A|}\mathrm{Imp}(A_L)
  -\frac{|\mathcal{I}_{A_R}|}{|\mathcal{I}_A|}\mathrm{Imp}(A_R),
\]
equivalently minimizing the summed child sum of squares
$\sum_{A'\in\{A_L,A_R\}}\sum_{s\in\mathcal{I}_{A'}}\lVert y_{s+h}-\bar{y}_{A'}\rVert_2^{2}$. Because the
impurity sums over maturities, one partition serves all maturities at once, which is what keeps the
residual cross-section coherent. Splitting continues until a stopping rule on the minimum node size and
maximum depth is met, and the tree's prediction at a query point $W$ is the node mean $\bar{y}_{A(W)}$
of the leaf $A(W)$ that contains it, so that $g_h^{(b)}(W)=\bar{y}_{A^{(b)}(W)}$ for tree $b$.

The trees are decorrelated in the two standard ways. Each tree $b$ is grown on a bootstrap resample
$\mathcal{I}_t^{(b)}$, drawn with replacement from $\mathcal{I}_t$, and at each node the splitting
coordinate is searched over a random subset of the $d_W$ predictors rather than all of them. The number
of trees $B$, the minimum node size, the maximum depth, the size of the random coordinate subset, and
whether resampling is used are chosen per window by randomized search with cross-validation. The full
rolling procedure is run under $10$ independent seeds, and the forecasts are averaged across seeds,
yielding the residual \eqref{eq:resid_rf}.

\subsection{Forecast evaluation}
\label{app:rmsfe}

Forecast accuracy at the one-month horizon is summarized by the maturity-specific root mean squared
forecast error,
\begin{equation}
  \mathrm{RMSFE}(\tau)=\Big(\frac{1}{n}\sum_{t} e^{\mathrm{b}}_{t+h}(\tau)^{2}\Big)^{1/2},
\label{eq:rmsfe}
\end{equation}
the sum running over the $n$ evaluation months, with $h=1$. Formal comparisons against the no-change
random walk are made not on this summary but on the squared-error loss differential through the
Diebold--Mariano statistic of Section~\ref{subsec:emp_forecast}, computed per specification by
Algorithm~\ref{alg:dmtest}.

\section{Forecast-combination baseline}
\label{app:combination}

This appendix defines the forecast-combination schemes used only as a downstream baseline in
Section~\ref{subsec:emp_baselines}. None enters the recovery framework, and none is a distributionally
robust combination; the notation here is local to this appendix. All eleven schemes take the same
inputs and differ only in how past forecast errors are turned into weights.

Combination is carried out separately for each maturity $\tau$ and horizon $h$, both of which we
suppress. Let $Y_t$ be the realized yield to be predicted at origin $t$ and let
$\hat{y}_{1,t},\ldots,\hat{y}_{M,t}$ be the $M$ candidate $h$-step forecasts, formed from the models
of Section~\ref{sec:benchmarks} and their variants. A combined forecast averages them with convex
weights,
\[
  \hat{y}^{c}_t=\sum_{m=1}^{M} w_{m,t}\,\hat{y}_{m,t},
  \qquad
  w_t=(w_{1,t},\ldots,w_{M,t})^{\top}\in\Delta_M
      =\Big\{w\in\mathbb{R}^{M}_{+}:\textstyle\sum_{m}w_m=1\Big\},
\]
with combined error $e^{c}_t=Y_t-\hat{y}^{c}_t$. Weights use only past errors. Over a rolling window
of $R=24$ months ending at $t-1$, each model's realized errors $e_{m,j}=Y_j-\hat{y}_{m,j}$,
$j=t-R,\ldots,t-1$, are collected into the error matrix
$\mathcal{E}=[e_{m,j}]\in\mathbb{R}^{R\times M}$, whose row $\mathcal{E}_j\in\mathbb{R}^{1\times M}$ is
the cross-model error vector at $j$ and whose $m$th column is model $m$'s error history. The rolling
accuracy of each model is $\widehat{\mathrm{RMSFE}}_{m,t}=(R^{-1}\sum_j e_{m,j}^{2})^{1/2}$ and
$\widehat{\mathrm{MSE}}_{m,t}=\widehat{\mathrm{RMSFE}}_{m,t}^{2}$.

\subsection{Accuracy-weighted schemes}
\label{app:comb_classic}
These assign weights from each model's past accuracy alone, ignoring correlation between models'
errors.
\begin{description}
\item[(1) FC--EW.] Equal weighting, $w_{m,t}=1/M$. It uses no past performance and serves as the
  benchmark the other schemes must beat.
\item[(2) FC--RANK.] Models are ordered by rolling accuracy and weighted by inverse rank. With
  $r_{m,t}\in\{1,\ldots,M\}$ the rank of $\widehat{\mathrm{RMSFE}}_{m,t}$ ($r=1$ most accurate),
  $w_{m,t}=r_{m,t}^{-1}/\sum_{k}r_{k,t}^{-1}$, so more accurate models receive more weight through a
  mapping insensitive to the magnitude of the accuracy gaps.
\item[(3) FC--RMSE.] Weight inversely proportional to rolling error,
  $w_{m,t}=\widehat{\mathrm{RMSFE}}_{m,t}^{-1}/\sum_{k}\widehat{\mathrm{RMSFE}}_{k,t}^{-1}$, a smooth
  version of FC--RANK that responds to the size of the differences.
\item[(4) FC--MSE.] Winner-take-all: all weight on the single most accurate model,
  $w_{m,t}=\mathbf{1}\{m=\arg\min_{k}\widehat{\mathrm{MSE}}_{k,t}\}$.
\item[(5) FC--OLS.] A regression combination in the spirit of \citet{granger1984improved}. The
  $q=\lceil 0.3M\rceil$ models with the lowest rolling RMSE are screened in; writing
  $\mathcal{E}_{(q),j}$ for the retained models' errors at $j$ and
  $\bar{e}_j=M^{-1}\sum_{m}e_{m,j}$ for the cross-model mean error, the fit
  $\min_{b}R^{-1}\sum_{j}(\bar{e}_j-\mathcal{E}_{(q),j}\,b)^{2}$ gives coefficients $b$ that are
  turned into weights by $w_{m,t}=|b_m|/\sum_{k}|b_k|$ for retained models and $0$ otherwise.
\end{description}

\subsection{Second-moment schemes}
\label{app:comb_var}
These use the full $M\times M$ error second-moment matrix, so a model is downweighted not only when
inaccurate but also when its errors are redundant with those of other models.
\begin{description}
\item[(6) FC--MV.] The minimum-variance combination of \citet{BatesGranger1969}: choose weights
  minimizing the variance of the combined error,
  $\min_{w\in\Delta_M} w^{\top}\hat{\Sigma}_\nu w$, where
  $\hat{\Sigma}_\nu=\mathrm{Cov}(\mathcal{E})+\nu I_M$ is the sample covariance of the error columns
  ridged by $\nu=10^{-6}$ for invertibility. The unconstrained optimum is
  $\hat{\Sigma}_\nu^{-1}\mathbf{1}/(\mathbf{1}^{\top}\hat{\Sigma}_\nu^{-1}\mathbf{1})$; negative
  weights are then set to zero and the remainder renormalized on $\Delta_M$.
\item[(7) FC--STACK.] Stacking on the uncentered second moment
  $\hat{S}=R^{-1}\mathcal{E}^{\top}\mathcal{E}$, so that squared bias as well as variance is
  penalized (FC--MV centers the errors and penalizes variance only):
  $\min_{w\in\Delta_M} w^{\top}\hat{S}w$, solved by projected gradient descent on the simplex.
\item[(8) FC--LAD.] A robust combination that replaces squared error by absolute error, limiting the
  influence of outlier months: $\min_{w\in\Delta_M} R^{-1}\sum_{j}\lvert\mathcal{E}_j w\rvert$, in the
  spirit of \citet{jiang2025robust}, solved approximately by iteratively reweighted least squares:
  the observation weights are $1/\max(\lvert\mathcal{E}_j w\rvert,10^{-6})$, the reweighted normal
  matrix is ridged by $\phi\cdot10^{-3}$ with $\phi=0.02$ for stability, and each step is projected
  onto the simplex.
\end{description}

\subsection{Exponentially reweighted schemes}
\label{app:comb_after}
The aggregated forecast through exponential reweighting of \citet{yang2004combining} downweights a
model geometrically in its cumulative squared error over the same trailing window of $R=24$ months as
the other schemes. The weights take the closed form
\[
  w_{m,t}\;\propto\;\hat{v}_{m,t}^{-1/2}\,
  \exp\!\Big(-\tfrac{1}{2\hat{v}_{m,t}}\sum_{j=t-R}^{t-1} e_{m,j}^{2}\Big),
\]
normalized to sum to one, where $\hat{v}_{m,t}>0$ is a variance estimate that both scales the
exponent and enters the leading factor, and is truncated below by a small constant for stability.
The three schemes differ only in $\hat{v}_{m,t}$.
\begin{description}
\item[(9) AFTER--Rolling.] Rolling sample variance of the model's errors over the window,
  $\hat{v}_{m,t}=\mathrm{Var}(e_{m,t-R},\ldots,e_{m,t-1})$.
\item[(10) AFTER--EWMA.] Exponentially weighted variance of the same errors with span $S=20$
  (decay $\delta=(S-1)/(S+1)=19/21$), giving recent months more weight in $\hat{v}_{m,t}$.
\item[(11) AFTER--Simplified.] The homoskedastic case $\hat{v}_{m,t}\equiv 1$, so weights reduce to
  pure exponential reweighting by cumulative squared error,
  $w_{m,t}\propto\exp\!\big(-\tfrac12\sum_{j=t-R}^{t-1} e_{m,j}^{2}\big)$.
\end{description}

\section{Tables}
\label{app:tables}

\begin{table}[H]
\centering
\caption{Explained variance and cumulative explained variance of the first ten principal components.}
\label{tab:cumulative_variance}
\small
\setlength{\tabcolsep}{8pt}
\renewcommand{\arraystretch}{1.1}
\begin{tabular}{lcc@{\hspace{2.5em}}lcc}
\toprule
PC & Explained & Cumulative & PC & Explained & Cumulative \\
\midrule
PC1 & 0.2304 & 0.2304 & PC6  & 0.0381 & 0.6172 \\
PC2 & 0.1180 & 0.3484 & PC7  & 0.0372 & 0.6544 \\
PC3 & 0.0977 & 0.4461 & PC8  & 0.0270 & 0.6814 \\
PC4 & 0.0807 & 0.5267 & PC9  & 0.0226 & 0.7040 \\
PC5 & 0.0523 & 0.5790 & PC10 & 0.0211 & 0.7251 \\
\bottomrule
\end{tabular}
\end{table}

\begin{sidewaystable}
\scriptsize
\centering
\caption{Categorized list of the $111$ macroeconomic, Treasury supply-demand, and financial
indicators in the predictor panel $Z_t$.}
\label{tab:EC_us_indicators}
\begin{tabular}{>{\raggedright\arraybackslash}p{4cm} >{\raggedright\arraybackslash}p{\dimexpr\textheight-5.5cm\relax}}
\toprule
\textbf{Category} & \textbf{Indicators} \\
\midrule
\textbf{Prices and Inflation} &
Consumer Price Index (All Urban, All Items), CPI Excluding Food and Energy (Core), CPI Annual
Inflation Rate, Producer Price Index (Final Demand), PPI Final Demand Excluding Foods and Energy, PPI
Final Demand (not seasonally adjusted), Chain-Type PCE Price Index, Chain-Type PCE Price Index
Excluding Food and Energy (Core), Export Price Index (All Commodities, End Use), Import Price Index
(All Commodities, End Use), GDP Implicit Price Deflator, Chain-Type Price Index of GDP \\
\midrule
\textbf{Labor Markets} &
Nonfarm Payroll Employment (Total), Total Civilian Employment, Unemployed Persons (16 Years and
Over), Unemployment Rate, Job Openings, Total Nonfarm (JOLTS), Average Hourly Earnings, Production
Employees (Total Private), Average Hourly Earnings, All Employees (Total Private), Average Hourly
Earnings, Production Employees (Manufacturing), Average Weekly Hours, Production Employees (Total
Private), Challenger Job-Cut Announcements, Employment Cost Index (Civilian Workers), Output per Hour
(Business Sector), Output per Hour (Nonfarm Business Sector), Unit Labor Costs (Business Sector),
Unit Labor Costs (Nonfarm Business Sector), Business Bankruptcy Filings (12-Month Total) \\
\midrule
\textbf{Real Activity} &
Industrial Production (Total Index), Industrial Production (Manufacturing), Capacity Utilization Rate
(All Industry), Retail Sales and Food Services (Total), Retail Sales and Food Services Excluding
Motor Vehicles and Parts, New Orders (All Manufacturing Industries), New Orders (Manufacturing
Excluding Transportation), New Orders (Manufacturing Durables), New Orders (Manufacturing Durables
Excluding Transportation), Business Inventories (Manufacturing and Trade), Business Sales
(Manufacturing and Trade), Wholesale Trade Inventories (Total), Sales by Merchant Wholesalers
(Total), Construction Expenditures (Total), New Passenger Car Registrations \\
\midrule
\textbf{National Accounts} &
Gross Domestic Product (current dollars), Gross Domestic Product (chained dollars), Gross National
Product (current dollars), Gross National Product (chained dollars), Personal Consumption
Expenditures, NIA (current dollars), Personal Consumption Expenditures, NIA (chained dollars),
Government Consumption and Investment (current dollars), Government Consumption and Investment
(chained dollars), Private Domestic Fixed Investment (current dollars), Private Domestic Fixed
Investment (chained dollars), Change in Private Inventories, Change in Private Inventories (NIA,
chained dollars), Corporate Profits (with IVA and CCAdj) \\
\midrule
\textbf{Business Conditions, Surveys, and Leading Indicators} &
Conference Board Leading Economic Indicators Index, ISM Manufacturing PMI, ISM Non-Manufacturing
Composite Index, Chicago Purchasing Managers Business Barometer, Philadelphia Fed Manufacturing
Business Outlook (General Activity), Empire State Manufacturing Survey (General Business Conditions),
TIPP Economic Optimism Index \\
\midrule
\textbf{Household and Personal Sector} &
Personal Income (Monthly Series), Disposable Personal Income, Personal Consumption Expenditures
(Monthly Series), Personal Saving Rate (Percent of Disposable Income), Consumer Credit Outstanding,
Consumer Confidence Index (Conference Board), Population (National-Accounts Estimate) \\
\midrule
\textbf{Housing Market} &
Housing Starts (New Private Units), New Private Housing Units Authorized by Building Permit, Housing
Units Authorized (unadjusted), New Single-Family Home Sales, Existing Home Sales (Single-Family and
Condominiums), Pending Home Sales Index, NAHB Housing Market Index \\
\midrule
\textbf{External Sector} &
Goods Exports (Balance-of-Payments Basis), Goods Imports (Balance-of-Payments Basis), Goods Trade
Balance (Balance-of-Payments Basis), Goods and Services Balance (Balance-of-Payments Basis), Exports
F.A.S., Imports F.A.S., Visible Trade Balance (F.A.S.--F.A.S.), Exports of Goods and Services, NIA
(current dollars), Exports of Goods and Services, NIA (chained dollars), Imports of Goods and
Services, NIA (current dollars), Imports of Goods and Services, NIA (chained dollars), Current
Account Balance, Capital and Financial Account Balance, Dollar--Euro Exchange Rate, Nominal
Advanced-Foreign-Economies Dollar Index, Terms of Trade \\
\midrule
\textbf{Financial Conditions and Interest Rates} &
Federal Funds Target Rate (end of period), Effective Federal Funds Rate (monthly average), Discount
Window Primary Credit Rate, Treasury Bill Rate (3-Month), Interbank Rate (3-Month, London), Prime
Rate Charged by Banks, Treasury Constant-Maturity Yield (20-Year), Dow Jones Industrials Share Price
Index, Monetary Base, Money Supply M1, Money Supply M2, Commercial Bank Loans and Leases in Bank
Credit, Commercial Bank Commercial and Industrial Loans \\
\midrule
\textbf{Treasury Supply and Capital Flows} &
Public Debt Outstanding, Total Treasury Securities Outstanding, Federal Government Budget Balance,
Foreign Net Long-Term Flows in Securities, Foreign Reserve Assets \\
\bottomrule
\end{tabular}

\vspace{3pt}
\begin{minipage}{\dimexpr\textheight-2cm\relax}\footnotesize\raggedright
\emph{Note.} The $111$ series as they appear in the panel; current-dollar and chained-dollar versions
of the national-accounts aggregates enter as separate indicators, and the monthly personal-outlays
series are distinct from their NIA counterparts. Every predictor enters lagged by at least one month
(Section~\ref{sec:data}).
\end{minipage}
\end{sidewaystable}

\begin{sidewaystable}
\centering
\begin{threeparttable}
\caption{One-step ($h=1$) Diebold--Mariano statistics against the no-change random walk, by specification,
over $115$ months.}
\label{tab:dm_byspec}
\footnotesize\setlength{\tabcolsep}{3pt}
\begin{tabular}{lccccccccccccccc}
\toprule
Spec. & 3M & 6M & 1Y & 2Y & 3Y & 4Y & 5Y & 6Y & 7Y & 8Y & 9Y & 10Y & 15Y & 20Y & 30Y \\
\midrule
\multicolumn{16}{l}{\emph{Random forest, by seed}}\\
seed 1466 & $-0.89$ & $0.28$ & $0.46$ & $-4.07^{***}$ & $-4.48^{***}$ & $-4.23^{***}$ & $-4.53^{***}$ & $-5.85^{***}$ & $-5.63^{***}$ & $-5.75^{***}$ & $-5.85^{***}$ & $-4.88^{***}$ & $-4.26^{***}$ & $-4.17^{***}$ & $-4.27^{***}$ \\
seed 1860 & $1.72^{*}$ & $1.66$ & $0.43$ & $-0.86$ & $-2.24^{**}$ & $-4.75^{***}$ & $-2.26^{**}$ & $-2.45^{**}$ & $-3.96^{***}$ & $-4.16^{***}$ & $-3.88^{***}$ & $-4.08^{***}$ & $-4.88^{***}$ & $-4.79^{***}$ & $-3.96^{***}$ \\
seed 5426 & $1.34$ & $0.97$ & $0.14$ & $-2.75^{***}$ & $-1.65$ & $-4.23^{***}$ & $-2.60^{**}$ & $-2.37^{**}$ & $-3.75^{***}$ & $-4.89^{***}$ & $-4.20^{***}$ & $-3.91^{***}$ & $-4.94^{***}$ & $-4.42^{***}$ & $-4.54^{***}$ \\
seed 6191 & $-1.04$ & $-0.26$ & $-1.12$ & $-3.86^{***}$ & $-4.71^{***}$ & $-4.83^{***}$ & $-4.88^{***}$ & $-5.44^{***}$ & $-5.85^{***}$ & $-4.78^{***}$ & $-4.15^{***}$ & $-4.87^{***}$ & $-6.16^{***}$ & $-4.61^{***}$ & $-5.35^{***}$ \\
seed 6390 & $2.00^{**}$ & $1.87^{*}$ & $0.83$ & $-2.67^{***}$ & $-2.16^{**}$ & $-4.01^{***}$ & $-3.69^{***}$ & $-3.28^{***}$ & $-3.39^{***}$ & $-4.86^{***}$ & $-2.57^{**}$ & $-3.23^{***}$ & $-4.12^{***}$ & $-4.87^{***}$ & $-3.58^{***}$ \\
seed 6578 & $0.18$ & $0.43$ & $-4.14^{***}$ & $-4.16^{***}$ & $-5.03^{***}$ & $-5.56^{***}$ & $-5.90^{***}$ & $-3.87^{***}$ & $-5.46^{***}$ & $-5.94^{***}$ & $-5.66^{***}$ & $-5.33^{***}$ & $-5.55^{***}$ & $-5.98^{***}$ & $-5.30^{***}$ \\
seed 6734 & $0.58$ & $0.90$ & $-0.41$ & $-3.45^{***}$ & $-3.05^{***}$ & $-4.66^{***}$ & $-3.97^{***}$ & $-4.14^{***}$ & $-5.75^{***}$ & $-6.18^{***}$ & $-5.92^{***}$ & $-5.50^{***}$ & $-4.83^{***}$ & $-5.89^{***}$ & $-6.19^{***}$ \\
seed 7265 & $1.39$ & $0.74$ & $0.28$ & $-3.93^{***}$ & $-3.04^{***}$ & $-4.86^{***}$ & $-5.46^{***}$ & $-4.81^{***}$ & $-5.09^{***}$ & $-6.11^{***}$ & $-5.83^{***}$ & $-5.46^{***}$ & $-5.89^{***}$ & $-5.47^{***}$ & $-4.13^{***}$ \\
seed 8270 & $1.31$ & $0.98$ & $-1.15$ & $-2.88^{***}$ & -- & $-3.61^{***}$ & $-4.39^{***}$ & $-4.17^{***}$ & $-4.30^{***}$ & $-5.70^{***}$ & $-5.56^{***}$ & $-5.36^{***}$ & $-5.42^{***}$ & $-4.84^{***}$ & $-3.88^{***}$ \\
seed 9322 & $1.21$ & $1.15$ & $0.27$ & $-2.96^{***}$ & $-5.00^{***}$ & $-4.09^{***}$ & $-3.59^{***}$ & $-4.13^{***}$ & $-4.70^{***}$ & $-3.87^{***}$ & $-3.56^{***}$ & $-3.41^{***}$ & $-3.66^{***}$ & $-5.27^{***}$ & $-4.44^{***}$ \\
\midrule
\multicolumn{16}{l}{\emph{FADNS, by factor count}}\\
$k=1$ & $2.02^{**}$ & $0.37$ & $0.46$ & $3.44^{***}$ & $3.79^{***}$ & $3.51^{***}$ & $2.89^{***}$ & $2.37^{**}$ & $2.08^{**}$ & $2.06^{**}$ & $2.26^{**}$ & $2.79^{***}$ & $3.92^{***}$ & $3.96^{***}$ & $3.44^{***}$ \\
$k=2$ & $2.13^{**}$ & $0.50$ & $0.48$ & $3.33^{***}$ & $3.75^{***}$ & $3.54^{***}$ & $2.97^{***}$ & $2.51^{**}$ & $2.27^{**}$ & $2.29^{**}$ & $2.51^{**}$ & $3.04^{***}$ & $4.02^{***}$ & $4.04^{***}$ & $3.70^{***}$ \\
$k=3$ & $2.21^{**}$ & $0.61$ & $0.51$ & $3.34^{***}$ & $3.76^{***}$ & $3.57^{***}$ & $3.05^{***}$ & $2.65^{***}$ & $2.48^{**}$ & $2.54^{**}$ & $2.76^{***}$ & $3.28^{***}$ & $4.18^{***}$ & $4.20^{***}$ & $4.05^{***}$ \\
$k=4$ & $2.21^{**}$ & $0.49$ & $0.22$ & $2.98^{***}$ & $3.51^{***}$ & $3.37^{***}$ & $2.87^{***}$ & $2.46^{**}$ & $2.29^{**}$ & $2.37^{**}$ & $2.66^{***}$ & $3.32^{***}$ & $4.45^{***}$ & $4.48^{***}$ & $4.15^{***}$ \\
$k=5$ & $2.22^{**}$ & $0.56$ & $0.41$ & $3.01^{***}$ & $3.54^{***}$ & $3.44^{***}$ & $2.94^{***}$ & $2.54^{**}$ & $2.36^{**}$ & $2.40^{**}$ & $2.65^{***}$ & $3.26^{***}$ & $4.38^{***}$ & $4.45^{***}$ & $4.19^{***}$ \\
$k=6$ & $2.31^{**}$ & $0.62$ & $0.14$ & $2.85^{***}$ & $3.45^{***}$ & $3.38^{***}$ & $2.93^{***}$ & $2.58^{**}$ & $2.43^{**}$ & $2.46^{**}$ & $2.64^{***}$ & $3.14^{***}$ & $4.33^{***}$ & $4.24^{***}$ & $4.30^{***}$ \\
$k=7$ & $2.51^{**}$ & $0.87$ & $0.10$ & $2.83^{***}$ & $3.44^{***}$ & $3.37^{***}$ & $2.91^{***}$ & $2.57^{**}$ & $2.44^{**}$ & $2.49^{**}$ & $2.71^{***}$ & $3.24^{***}$ & $4.47^{***}$ & $4.45^{***}$ & $4.47^{***}$ \\
$k=8$ & $2.42^{**}$ & $0.86$ & $0.34$ & $2.94^{***}$ & $3.48^{***}$ & $3.38^{***}$ & $2.98^{***}$ & $2.67^{***}$ & $2.58^{**}$ & $2.70^{***}$ & $2.99^{***}$ & $3.52^{***}$ & $4.64^{***}$ & $4.62^{***}$ & $4.46^{***}$ \\
$k=9$ & $2.38^{**}$ & $0.94$ & $0.45$ & $2.90^{***}$ & $3.44^{***}$ & $3.35^{***}$ & $2.95^{***}$ & $2.65^{***}$ & $2.56^{**}$ & $2.67^{***}$ & $2.92^{***}$ & $3.42^{***}$ & $4.45^{***}$ & $4.41^{***}$ & $4.32^{***}$ \\
$k=10$ & $2.61^{**}$ & $1.12$ & $0.55$ & $2.91^{***}$ & $3.47^{***}$ & $3.40^{***}$ & $3.01^{***}$ & $2.72^{***}$ & $2.64^{***}$ & $2.73^{***}$ & $2.96^{***}$ & $3.44^{***}$ & $4.40^{***}$ & $4.39^{***}$ & $4.48^{***}$ \\
\bottomrule
\end{tabular}
\begin{tablenotes}[flushleft]\footnotesize
\item Entries are the Diebold--Mariano statistic $\mathrm{DM}(\tau)$ of
Section~\ref{subsec:emp_forecast}; a negative value rejects in the specification's favor, a positive value
against it ($^{*}p<0.10$, $^{**}p<0.05$, $^{***}p<0.01$). A dash marks a maturity absent for that seed. The
column medians and ranges give the cells of Table~\ref{tab:forecast_summary}.
\end{tablenotes}
\end{threeparttable}
\end{sidewaystable}

\begin{table}[H]
\centering
\begin{threeparttable}
\caption{Selected indicators and coefficients by specification.}
\label{tab:naming_full}
\footnotesize
\begin{tabular}{lp{0.80\textwidth}}
\toprule
Spec. & Selected indicators (coefficient) \\
\midrule
RF & M2 money supply ($-0.24$); Chicago PMI ($-0.09$); building permits ($-0.18$); change in private
inventories ($-1.01$) \\
$k=1$ & Conference Board leading index ($+1.02$); dollar--euro rate ($-0.07$); advanced-economies dollar
index ($+0.08$); goods and services balance ($-0.06$) \\
$k=2$ & change in private inventories ($+0.15$) \\
$k=3$ & Conference Board leading index ($+0.93$); dollar--euro rate ($-0.09$); change in private
inventories ($+0.04$) \\
$k=4$ & Conference Board leading index ($-0.96$); M2 money supply ($-1.20$); consumer confidence
($+0.78$); NAHB housing market index ($+0.72$); Chicago PMI ($+0.21$); federal funds target rate
($+2.40$); Treasury securities outstanding ($+1.50$); foreign net long-term securities flows ($-0.87$);
average weekly hours, production employees ($+1.04$); Challenger job-cut announcements ($-0.81$); new
single-family home sales ($-0.28$); industrial production ($-0.48$); building permits ($+0.14$);
construction expenditures ($+4.82$); goods trade balance ($-0.96$); change in private inventories
($+0.10$); change in private inventories, NIA ($+0.00$); capital and financial account balance
($-0.10$) \\
$k=5$ & Conference Board leading index ($-0.86$); M2 money supply ($-1.15$); consumer confidence
($+0.87$); NAHB housing market index ($+0.89$); Chicago PMI ($+0.23$); federal funds target rate
($+2.40$); Treasury securities outstanding ($+1.93$); foreign net long-term securities flows ($-0.91$);
average weekly hours, production employees ($+1.07$); Challenger job-cut announcements ($-0.84$); new
single-family home sales ($-0.39$); industrial production ($-0.40$); building permits ($+0.04$);
construction expenditures ($+3.86$); goods exports ($-1.47$); goods trade balance ($-0.80$); government
consumption and investment ($-0.14$); change in private inventories ($+0.44$); capital and financial
account balance ($-0.02$) \\
$k=6$ & Conference Board leading index ($-1.59$); federal funds target rate ($+2.14$); foreign net
long-term securities flows ($-0.55$); Challenger job-cut announcements ($-1.09$); industrial production
($-0.12$); building permits ($+0.92$); federal budget balance ($-0.02$); goods trade balance ($-0.12$);
goods and services balance ($-0.08$) \\
$k=7$ & Conference Board leading index ($-1.70$); federal funds target rate ($+2.27$); foreign net
long-term securities flows ($-0.89$); Challenger job-cut announcements ($-1.29$); industrial production
($-0.15$); building permits ($+0.87$); federal budget balance ($-0.01$); goods trade balance ($-0.87$) \\
$k=8$ & Conference Board leading index ($-1.70$); federal funds target rate ($+2.16$); foreign net
long-term securities flows ($-0.87$); Challenger job-cut announcements ($-1.29$); industrial production
($-0.07$); building permits ($+0.64$); goods trade balance ($-0.85$) \\
$k=9$ & Conference Board leading index ($-2.03$); consumer confidence ($+0.84$); monetary base
($+0.01$); three-month interbank rate ($+1.34$); foreign net long-term securities flows ($-0.47$);
Challenger job-cut announcements ($-0.76$); industrial production ($-0.00$); building permits ($+0.51$);
goods trade balance ($-0.14$); goods and services balance ($-0.07$) \\
$k=10$ & Conference Board leading index ($-1.81$); consumer confidence ($+0.43$); Chicago PMI ($+0.05$);
three-month interbank rate ($+1.58$); foreign net long-term securities flows ($-0.43$); Challenger
job-cut announcements ($-0.73$); building permits ($+0.20$); goods trade balance ($-0.15$); goods and
services balance ($-0.09$); housing authorized ($+0.16$) \\
\bottomrule
\end{tabular}
\begin{tablenotes}[flushleft]\footnotesize
\item Indicators selected by the naming regression of Table~\ref{tab:naming}, with coefficients in
standardized units. The overall sign of each specification's factor, hence of its coefficient vector,
is arbitrary.
\end{tablenotes}
\end{threeparttable}
\end{table}

\begin{table}[H]
\centering
\begin{threeparttable}
\caption{Linear-against-forest contrast of selected drivers and recovered directions.}
\label{tab:contrast}
\begin{tabular}{lcccc}
\toprule
Spec. & L only & RF only & Both & Angle (deg) \\
\midrule
$k=1$  & $4$  & $4$ & $0$ & $38.8$ \\
$k=2$  & $0$  & $3$ & $1$ & $42.4$ \\
$k=3$  & $2$  & $3$ & $1$ & $48.7$ \\
$k=4$  & $14$ & $0$ & $4$ & $72.2$ \\
$k=5$  & $15$ & $0$ & $4$ & $72.6$ \\
$k=6$  & $8$  & $3$ & $1$ & $66.7$ \\
$k=7$  & $7$  & $3$ & $1$ & $67.2$ \\
$k=8$  & $6$  & $3$ & $1$ & $65.5$ \\
$k=9$  & $9$  & $3$ & $1$ & $65.8$ \\
$k=10$ & $8$  & $2$ & $2$ & $65.4$ \\
\bottomrule
\end{tabular}
\begin{tablenotes}[flushleft]\footnotesize
\item For each factor count $k$: indicators selected only by the FADNS specification, only by the
forest, and by both, and the angle between the averaged recovered directions $v^{\mathrm{L}}$ and
$v^{\mathrm{RF}}$.
\end{tablenotes}
\end{threeparttable}
\end{table}

\section{Algorithms}
\label{app:algorithms}

\begin{algorithm}[H]
\caption{Rolling Diebold--Li dynamic Nelson--Siegel (DNS) forecasting}
\label{alg:dns}
\begin{algorithmic}[1]
\Require monthly zero-coupon yields $\{y_t(\tau_j)\}_{j=1}^{N}$; window $w=60$; horizon $h=1$;
         decay $\lambda=0.0609$
\Ensure one-step yield forecasts and forecast errors
\For{$t=w$ \textbf{to} $T-h$}
  \For{$s=t-w+1$ \textbf{to} $t$}
    \State fit DNS factors $\beta_s=(\beta_{1s},\beta_{2s},\beta_{3s})^{\top}$ by cross-sectional
           least squares \Comment{fixed loadings, \eqref{eq:dns}}
  \EndFor
  \State fit VAR(1) $\beta_{s+1}=c+\Phi\beta_s+\eta_s$ on the window
  \State $\hat{\beta}_{t+h\mid t}\gets c+\Phi\beta_t$ \Comment{one-step conditional mean, $h=1$}
  \For{\textbf{each} maturity $\tau_j$}
    \State $\hat{y}^{\mathrm{L}}_{t+h\mid t}(\tau_j)\gets$ Nelson--Siegel map of $\hat{\beta}_{t+h\mid t}$;\quad
           store $e^{\mathrm{L}}_{t+h}(\tau_j)=y_{t+h}(\tau_j)-\hat{y}^{\mathrm{L}}_{t+h\mid t}(\tau_j)$
  \EndFor
\EndFor
\end{algorithmic}
\end{algorithm}

\begin{algorithm}[H]
\caption{Rolling factor-augmented dynamic Nelson--Siegel (FADNS) forecasting}
\label{alg:fadns}
\begin{algorithmic}[1]
\Require yields $\{y_t(\tau_j)\}_{j=1}^{N}$; predictor panel $\{Z_t\in\mathbb{R}^{p}\}$, $p=111$;
         window $w=60$; horizon $h=1$; factor counts $k\in\{1,\dots,10\}$; decay $\lambda=0.0609$
\Ensure one-step forecasts and errors for every factor count $k$
\For{\textbf{each} $k\in\{1,\dots,10\}$} \Comment{all ten specifications are carried; none is selected}
  \For{$t=w$ \textbf{to} $T-h$}
    \For{$s=t-w+1$ \textbf{to} $t$}
      \State fit DNS factors $\beta_s$ by cross-sectional least squares
    \EndFor
    \State form the lagged block $\{Z_{t-w},\dots,Z_{t-1}\}$; unit-root filter (ADF, $10\%$) and
           standardize within the window \Comment{filter and scaling are $k$-free}
    \State PCA on the standardized block; component $j$ at $t$ uses information dated $t-1$ and
           earlier, $\mathrm{PC}_{j,t}=v_{j,t}^{\top}\tilde{Z}_{t-1}$;\quad
           $F^{(k)}_t\gets(\mathrm{PC}_{1,t},\dots,\mathrm{PC}_{k,t})^{\top}$
    \State $X^{(k)}_t\gets(\beta_t^{\top},F^{(k)\top}_t)^{\top}$;\quad
           fit VAR(1) $X^{(k)}_{s+1}=c^{(k)}+\Phi^{(k)}X^{(k)}_s+\eta^{(k)}_s$ on the window
    \State $\hat{X}^{(k)}_{t+h\mid t}\gets c^{(k)}+\Phi^{(k)}X^{(k)}_t$;\quad extract
           $\hat{\beta}_{t+h\mid t}$;\quad Nelson--Siegel map;\quad store
           $e^{\mathrm{L},k}_{t+h}(\tau_j)$ \Comment{\eqref{eq:resid_fadns}}
  \EndFor
\EndFor
\end{algorithmic}
\end{algorithm}

\begin{algorithm}[H]
\caption{Rolling joint random-forest forecasting}
\label{alg:rf}
\begin{algorithmic}[1]
\Require yield vectors $\{y_t\in\mathbb{R}^{N}\}_{t=1}^{T}$; predictor panel $\{Z_t\}$; horizon $h=1$;
         window $w=60$; seeds $\mathcal{S}$, $|\mathcal{S}|=10$; hyperparameter space $\Theta$
\Ensure per-seed errors and the seed-averaged forecast
\For{\textbf{each} seed $s\in\mathcal{S}$}
  \For{\textbf{each} origin $t=w,\dots,T-h$}
    \State $\check{W}_t\gets\big((Z_{t-\ell})_{\ell\in\mathcal{L}_Z},(y_{t-\ell})_{\ell\in\mathcal{L}_y}\big)$,
           $\mathcal{L}_Z=\{1,\dots,60\}$, $\mathcal{L}_y=\{0,\dots,59\}$ \Comment{\eqref{eq:rf_predictor}}
    \State training sample $\mathcal{D}_t=\{(\check{W}_u,\,y_{u+h}):u=t-w+1,\dots,t\}$;\quad
           min--max scale predictors over the window
    \State $\theta^{\ast}\gets\arg\min_{\theta\in\Theta}\mathrm{CV\text{-}MSE}(\theta;\mathcal{D}_t)$
           by randomized cross-validation
    \State fit a single joint multi-output forest $g_h^{(s)}$ on $\mathcal{D}_t$ with $\theta^{\ast}$;
           store $e^{(s)}_{t+h}=y_{t+h}-g_h^{(s)}(W_t)$ \Comment{\eqref{eq:rf}, \eqref{eq:rf_impurity}}
  \EndFor
\EndFor
\State $\hat{y}^{\mathrm{RF}}_{t+h\mid t}\gets|\mathcal{S}|^{-1}\sum_{s}g_h^{(s)}(W_t)$;\quad
       $e^{\mathrm{RF}}_{t+h}=y_{t+h}-\hat{y}^{\mathrm{RF}}_{t+h\mid t}$
       \Comment{seed average for recovery, \eqref{eq:resid_rf}; per-seed errors retained for ranges}
\end{algorithmic}
\end{algorithm}

\begin{algorithm}[H]
\caption{Expanding residual-based DRO factor recovery for benchmark $\mathrm{b}$}
\label{alg:recovery}
\begin{algorithmic}[1]
\Require horizon $h=1$; burn-in $M_0=36$; residuals
         $e^{\mathrm{b}}_s=y_s-\hat{y}^{\mathrm{b}}_{s\mid s-h}$ from the fixed forecaster;
         forecast-implied mean $\hat{\mu}^{\mathrm{b}}_t=y_t-\hat{y}^{\mathrm{b}}_{t+h\mid t}$
         \Comment{$=-\widehat{\Delta y}$, \eqref{eq:mean_input}}
\Ensure recovered directions $\{v^{\mathrm{b}}_t\}$, out-of-sample factors
        $\{\xi^{\mathrm{b}}_{t+h}\}$, loadings $\{\kappa^{\mathrm{b}}_t\}$
\For{$t=M_0,\dots,T-h$} \Comment{training set is all months up to $t$}
  \State $\bar{e}^{\mathrm{b}}_t\gets$ mean of $\{e^{\mathrm{b}}_s\}_{s\le t}$;\quad
         $D^{\mathrm{b}}_t\gets\operatorname{diag}$ of per-maturity std.\ devs.\ over $s\le t$
         \Comment{\eqref{eq:vol_scaling}}
  \State $z^{\mathrm{b}}_s\gets(D^{\mathrm{b}}_t)^{-1}(e^{\mathrm{b}}_s-\bar{e}^{\mathrm{b}}_t)$,
         $s\le t$;\quad $\hat{\Sigma}^{\mathrm{b}}_{z,t}\gets$ Ledoit--Wolf shrinkage
         \Comment{\eqref{eq:std_residual}--\eqref{eq:shrunk_cov}}
  \State $\omega^{\mathrm{b},\ast}_t(0)\gets\big(D^{\mathrm{b}}_t\hat{\Sigma}^{\mathrm{b}}_{z,t}D^{\mathrm{b}}_t\big)^{-1}\hat{\mu}^{\mathrm{b}}_t$
         \Comment{closed-form tangency; Prop.~\ref{prop:robust_reduction} at $\rho=0$}
  \State $a^{\mathrm{b}}_t\gets D^{\mathrm{b}}_t\omega^{\mathrm{b},\ast}_t(0)$
  \State $\hat{d}^{\mathrm{b}}_t\gets\big(I_N-\tfrac{a^{\mathrm{b}}_t(a^{\mathrm{b}}_t)^{\top}}{\|a^{\mathrm{b}}_t\|^{2}}\big)\hat{\Sigma}^{\mathrm{b}}_{z,t}a^{\mathrm{b}}_t$;\quad
         $v^{\mathrm{b}}_t\gets\hat{d}^{\mathrm{b}}_t/\|\hat{d}^{\mathrm{b}}_t\|$;\quad
         $\kappa^{\mathrm{b}}_t\gets\|\hat{d}^{\mathrm{b}}_t\|$
         \Comment{\eqref{eq:forcing}, \eqref{eq:recovered_loading}}
  \State $z^{\mathrm{b}}_{t+h}\gets(D^{\mathrm{b}}_t)^{-1}(e^{\mathrm{b}}_{t+h}-\bar{e}^{\mathrm{b}}_t)$;\quad
         $\xi^{\mathrm{b}}_{t+h}\gets(v^{\mathrm{b}}_t)^{\top}z^{\mathrm{b}}_{t+h}$
         \Comment{standardized with training moments only; out of sample}
\EndFor
\Statex \textbf{Remarks.} At $\rho=0$ the worst-case Sharpe optimizer is the closed-form tangency
        direction, so no numerical program is solved. For $\gamma_1>0$ and $\rho>0$ the exposure
        solves \eqref{eq:robust_sharpe_reduced} in normalized form, the maximization of
        $m_t^{\top}a-\sqrt{\gamma_{1,t}}\,\lVert a\rVert_{\hat{\Sigma}^{\mathrm{b}}_{z,t}}$ under
        $\lVert a\rVert^{2}_{\hat{\Sigma}^{\mathrm{b}}_{z,t}+\rho_t I}\le1$, a concave program with one
        convex quadratic constraint (Algorithm~\ref{alg:decision}).
        When no mean input is formed, replace line~3 by
        $\omega\propto(D^{\mathrm{b}}_t\hat{\Sigma}^{\mathrm{b}}_{z,t}D^{\mathrm{b}}_t)^{-1}\mathbf{1}$
        (minimum variance) or $\omega\propto\mathbf{1}$ (equal weight); lines~4--6 are unchanged.
\end{algorithmic}
\end{algorithm}

\begin{algorithm}[H]
\caption{Expanding deployment of the duration decision with data-driven radii}
\label{alg:decision}
\begin{algorithmic}[1]
\Require residuals and mean input as in Algorithm~\ref{alg:recovery}; burn-in $M_0=36$;
         confidence $\delta=0.05$; support quantile $q=0.95$; rules
         $\{\text{single},\text{combination},\text{one-layer},\text{two-layer}\}$
\Ensure realized profit-and-loss series and Sharpe ratio per rule
\For{$t=M_0,\dots,T-h$}
  \State standardize and shrink on $s\le t$ as in Algorithm~\ref{alg:recovery};\quad
         $M_t\gets t$;\quad $m_t\gets(D^{\mathrm{b}}_t)^{-1}\hat{\mu}^{\mathrm{b}}_t$
  \State $R^2_t\gets$ $q$-quantile of
         $\{z_s^{\top}(\hat{\Sigma}^{\mathrm{b}}_{z,t})^{-1}z_s\}_{s\le t}$;\quad
         $\gamma_{1,t}\gets\tfrac{R^2_t}{M_t}\big(2+\sqrt{2\ln(1/\delta)}\big)^2$;\quad
         $\rho_t\gets\operatorname{tr}(\hat{\Sigma}^{\mathrm{b}}_{z,t})/M_t$
         \Comment{\eqref{eq:gamma1_calib}--\eqref{eq:rho_calib}}
  \State \textbf{single:} $a\propto(\hat{\Sigma}^{\mathrm{b}}_{z,t})^{-1}m_t$
         \Comment{$\gamma_1=\rho=0$; own forecast}
  \State \textbf{combination:} as single with $m_t$ from the pool-combined forecast
         \Comment{Algorithms~\ref{alg:fc_classical}--\ref{alg:fc_after}; pools never crossed}
  \State \textbf{one-layer:} if
         $\lVert m_t\rVert_{(\hat{\Sigma}^{\mathrm{b}}_{z,t})^{-1}}>\sqrt{\gamma_{1,t}}$ then
         $a\propto(\hat{\Sigma}^{\mathrm{b}}_{z,t})^{-1}m_t$, else $a\gets0$
         \Comment{optimal value $\lVert m_t\rVert_{\Sigma^{-1}}-\sqrt{\gamma_{1,t}}$ at $\rho=0$}
  \State \textbf{two-layer:} $a\gets\argmax_{\lVert a\rVert^{2}_{\hat{\Sigma}^{\mathrm{b}}_{z,t}+\rho_t I}\le1}
         \big(m_t^{\top}a-\sqrt{\gamma_{1,t}}\,\lVert a\rVert_{\hat{\Sigma}^{\mathrm{b}}_{z,t}}\big)$;
         if the optimum is $\le0$, $a\gets0$
         \Comment{normalized form of \eqref{eq:robust_sharpe_reduced}; withdrawn on a nonpositive optimum}
  \State scale each nonzero $a$ to $a^{\top}\hat{\Sigma}^{\mathrm{b}}_{z,t}a=1$;\quad
         $\Pi_{t+h}\gets-\,a^{\top}(D^{\mathrm{b}}_t)^{-1}\Delta y_{t+h}$
         \Comment{unit ex-ante volatility, \eqref{eq:pnl}}
\EndFor
\State report the realized Sharpe ratio of $\{\Pi_{t+h}\}$ per rule
\end{algorithmic}
\end{algorithm}

\begin{algorithm}[H]
\caption{Factor-neutral hedge of the two-layer position}
\label{alg:hedge}
\begin{algorithmic}[1]
\Require two-layer exposure $a_t$ at unit ex-ante volatility (Algorithm~\ref{alg:decision});
         direction $v^{\mathrm{b}}_t$ and loading $\kappa^{\mathrm{b}}_t$ (Algorithm~\ref{alg:recovery})
\Ensure profit-and-loss series of the hedged variants; realized volatility, maximum drawdown,
        $\mathrm{CVaR}_{5\%}$
\For{\textbf{each} origin $t$ with $a_t\neq0$}
  \State $g_t\gets\hat{\Sigma}^{\mathrm{b}}_{z,t}v^{\mathrm{b}}_t$;\quad
         $a^{\mathrm{h}}_t\gets a_t-\dfrac{g_t^{\top}a_t}{\lVert g_t\rVert^{2}}\,g_t$
         \Comment{projection onto \eqref{eq:hedge_constraint}; sets $\kappa=0$}
  \State \textbf{fixed budget:} deploy $a^{\mathrm{h}}_t$
         \Comment{removal without the rescaling it induces}
  \State \textbf{renormalized:} deploy
         $a^{\mathrm{h}}_t\big/\big((a^{\mathrm{h}}_t)^{\top}\hat{\Sigma}^{\mathrm{b}}_{z,t}a^{\mathrm{h}}_t\big)^{1/2}$
         \Comment{restore unit ex-ante volatility}
  \State $\Pi_{t+h}\gets-\,a^{\top}(D^{\mathrm{b}}_t)^{-1}\Delta y_{t+h}$ for each variant
\EndFor
\State report realized volatility, maximum drawdown of cumulative $\Pi$, and
       $\mathrm{CVaR}_{5\%}$, against the unhedged two-layer position
\end{algorithmic}
\end{algorithm}

\begin{algorithm}[H]
\caption{FarmSelect factor naming for benchmark $\mathrm{b}$}
\label{alg:naming}
\begin{algorithmic}[1]
\Require out-of-sample factors $\{\xi^{\mathrm{b}}_{t+h}\}$ (Algorithm~\ref{alg:recovery}) and
         standardized indicator vectors $\{x_{t+h}\}\subset\mathbb{R}^{p}$ over the $n$ naming
         origins; maximum factors $r_{\max}=8$; SCAD shape $a=3.7$ and a $\lambda$-grid; Huber
         constant $1.345$
\Ensure common-factor loadings $\hat{\alpha}$, idiosyncratic loadings $\hat{\theta}$, naming
        statistic $\hat{T}^{\mathrm{obs}}$
\State $X\gets[x_{t+h}^{\top}]$;\quad $\hat{\Sigma}_X\gets n^{-1}X^{\top}X$; eigenpairs
       $(\hat{\lambda}_j,\hat{q}_j)$, $\hat{\lambda}_1\ge\cdots\ge\hat{\lambda}_p$
\State $\hat{r}\gets\argmax_{1\le j\le r_{\max}}\hat{\lambda}_j/\hat{\lambda}_{j+1}$;\quad
       $\hat{B}\gets[\sqrt{\hat{\lambda}_1}\hat{q}_1,\dots,\sqrt{\hat{\lambda}_{\hat{r}}}\hat{q}_{\hat{r}}]$
       \Comment{eigenvalue-ratio rule}
\State $\hat{f}_{t+h}\gets(\hat{B}^{\top}\hat{B})^{-1}\hat{B}^{\top}x_{t+h}$;\quad
       $\hat{u}_{t+h}\gets x_{t+h}-\hat{B}\hat{f}_{t+h}$ \Comment{decorrelation, \eqref{eq:decorrelated}}
\For{\textbf{each} $\lambda$ on the grid}
  \State fit \eqref{eq:farm_obj}: minimize Huber loss $+$ SCAD$(\theta)$ with $\alpha$ unpenalized,
         iterating the robust scale $\hat{s}$ and threshold $\delta=1.345\,\hat{s}$
\EndFor
\State choose $\lambda$ by ten-fold cross-validation, the factor extraction repeated within every
       training fold and held-out rows projected through the training-fold loadings
       $\Rightarrow(\hat{\alpha},\hat{\theta})$
\State $\hat{T}^{\mathrm{obs}}\gets$ the cross-validated out-of-sample $R^{2}$ at the selected
       $\lambda$, the maximum over the grid
\end{algorithmic}
\end{algorithm}

\begin{algorithm}[H]
\caption{Selection-aware block-permutation $p$-value}
\label{alg:permtest}
\begin{algorithmic}[1]
\Require observed statistic $\hat{T}^{\mathrm{obs}}$ (Algorithm~\ref{alg:naming}); block length
         $\ell=\operatorname{round}(n^{1/3})$; draws $L=199$ with a fixed seed
\Ensure exact permutation $p$-value $\hat{p}$
\If{$\hat{T}^{\mathrm{obs}}\le 0$} \Return $\hat{p}\gets1$
      \Comment{skill gate: no positive out-of-sample fit to certify}
\EndIf
\State $c\gets 0$
\For{$q=1,\dots,L$}
  \State draw a block permutation (length $\ell$) of $\{\xi^{\mathrm{b}}_{t+h}\}$ and re-pair with $X$
  \State re-run Algorithm~\ref{alg:naming} on the permuted response, including the
         cross-validated choice of $\lambda$, $\Rightarrow\hat{T}^{(q)}$
         \Comment{nested selection: the null reflects the best fit achievable by chance}
  \State \textbf{if } $\hat{T}^{(q)}\ge\hat{T}^{\mathrm{obs}}$ \textbf{ then } $c\gets c+1$
\EndFor
\State $\hat{p}\gets(1+c)/(1+L)$ \Comment{fixed draws and seed; the $p$-value is reproducible}
\end{algorithmic}
\end{algorithm}

\begin{algorithm}[H]
\caption{Forecast combination schemes: accuracy-weighted and second-moment (1--8)}
\label{alg:fc_classical}
\begin{algorithmic}[1]
\Require notation as in Appendix~\ref{app:combination}: trailing forecast errors
         $\mathcal{E}=[e_{m,j}]\in\mathbb{R}^{R\times M}$, $j=t-R,\ldots,t-1$, over the rolling
         window $R=24$; the $M=10$ candidates are either the ten random-forest seeds or the ten
         FADNS factor counts, never crossed; per maturity, candidates lacking that maturity are
         dropped from the pool; ridge $\nu=10^{-6}$; LAD penalty $\phi=0.02$
\Ensure weights $w_t\in\Delta_M$
\State $\widehat{\mathrm{MSE}}_{m,t}=\tfrac{1}{R}\sum_{j=t-R}^{t-1}e_{m,j}^2$,\quad
       $\widehat{\mathrm{RMSFE}}_{m,t}=\widehat{\mathrm{MSE}}_{m,t}^{1/2}$

\vspace{1mm}\Statex \textbf{(1) FC--EW.}\quad $w_{m,t}=1/M$

\vspace{1mm}\Statex \textbf{(2) FC--RANK.}\quad
       $\tilde w_{m}\propto r_{m,t}^{-1}$, with $r_{m,t}$ the ascending rank of
       $\widehat{\mathrm{RMSFE}}_{m,t}$

\vspace{1mm}\Statex \textbf{(3) FC--RMSE.}\quad
       $\tilde w_{m}\propto 1/\max\big(\widehat{\mathrm{RMSFE}}_{m,t},10^{-8}\big)$

\vspace{1mm}\Statex \textbf{(4) FC--MSE.}\quad
       $w_{m,t}=\mathbf{1}\{m=\arg\min_k\widehat{\mathrm{MSE}}_{k,t}\}$

\vspace{1mm}\Statex \textbf{(5) FC--OLS.}
\State screen in the $q=\lceil 0.3M\rceil$ models of smallest $\widehat{\mathrm{RMSFE}}_{m,t}$;
       write $\mathcal{E}_{(q)}$ for the retained columns of $\mathcal{E}$ and
       $\bar{e}_j=M^{-1}\sum_{m}e_{m,j}$ for the cross-model mean error
\State fit $\min_{b}\tfrac{1}{R}\sum_{j}\big(\bar{e}_j-\mathcal{E}_{(q),j}\,b\big)^{2}$ without
       intercept;\quad $\tilde w_{m}=|b_m|$ for retained $m$,\; $\tilde w_{m}=0$ otherwise

\vspace{1mm}\Statex \textbf{(6) FC--MV (minimum variance).}
\State $\hat{\Sigma}_\nu=\mathrm{Cov}(\mathcal{E})+\nu I_M$;\quad
       $\tilde w=\hat{\Sigma}_\nu^{-1}\mathbf{1}/(\mathbf{1}^\top\hat{\Sigma}_\nu^{-1}\mathbf{1})$;\;
       clip $\tilde w\gets\max(\tilde w,0)$

\vspace{1mm}\Statex \textbf{(7) FC--STACK.}
\State $\hat{S}=\tfrac{1}{R}\mathcal{E}^\top\mathcal{E}$;\quad
       $\tilde w=\arg\min_{w}\tfrac12 w^\top \hat{S} w$ s.t.\ $\mathbf{1}^\top w=1,\;w\ge0$,
       by projected gradient descent on the simplex

\vspace{1mm}\Statex \textbf{(8) FC--LAD.}
\State solve $\min_{w\in\Delta_M}\tfrac{1}{R}\sum_{j}\lvert \mathcal{E}_j w\rvert$ approximately by
       iteratively reweighted least squares: observation weights
       $1/\max(\lvert \mathcal{E}_j w\rvert,10^{-6})$, reweighted normal matrix ridged by
       $\phi\cdot10^{-3}$, projected-gradient step onto the simplex \Comment{\citep{jiang2025robust}}

\vspace{1mm}
\State normalize $w_t=\tilde w/(\mathbf{1}^\top\tilde w)$;\quad
       if $\mathbf{1}^\top\tilde w=0$, set $w_t=\mathbf{1}/M$
\end{algorithmic}
\end{algorithm}

\begin{algorithm}[H]
\caption{Forecast combination schemes: exponentially reweighted (9--11)}
\label{alg:fc_after}
\begin{algorithmic}[1]
\Require trailing errors $\mathcal{E}$ as in Algorithm~\ref{alg:fc_classical}, over the same
         rolling window $R=24$; EWMA span $S=20$
\Ensure weights $w_t\in\Delta_M$
\State cumulative loss over the trailing window: $C_{m,t}=\sum_{j=t-R}^{t-1}e_{m,j}^2$
\Statex Following \citet{yang2004combining}, all three schemes take the closed form
       $\tilde w_{m}\propto \hat v_{m,t}^{-1/2}\exp\!\big(-C_{m,t}/(2\hat v_{m,t})\big)$
       and differ only in the variance estimate $\hat v_{m,t}$, floored at
       $\hat v_{m,t}\gets\max(\hat v_{m,t},10^{-6})$.

\vspace{1mm}\Statex \textbf{(9) AFTER--Rolling.}\quad
       $\hat v_{m,t}=\mathrm{Var}\big(e_{m,t-R},\dots,e_{m,t-1}\big)$

\vspace{1mm}\Statex \textbf{(10) AFTER--EWMA.}\quad
       $\hat v_{m,t}=$ EWMA variance of $\{e_{m,j}\}_{j=t-R}^{t-1}$, span $S=20$
       (decay $\delta=19/21$)

\vspace{1mm}\Statex \textbf{(11) AFTER--Simplified.}\quad homoskedastic $\hat v_{m,t}\equiv1$, so
       $\tilde w_{m}\propto\exp(-C_{m,t}/2)$

\vspace{1mm}
\State normalize $w_t=\tilde w/(\mathbf{1}^\top\tilde w)$;\quad
       if $\mathbf{1}^\top\tilde w=0$, set $w_t=\mathbf{1}/M$
\end{algorithmic}
\end{algorithm}

\begin{algorithm}[H]
\caption{Forecast-accuracy test against the no-change random walk}
\label{alg:dmtest}
\begin{algorithmic}[1]
\Require realized yields $\{y_t(\tau)\}$; one-step errors $\{e^{s}_{t+1}(\tau)\}$ for specifications
         $s=1,\dots,S$ ($S=10$ per forecaster), maturities $\tau$, over $n$ months; level $\alpha$
\Ensure per-maturity median $\mathrm{DM}$ statistic, range, and significance
\State $e^{\mathrm{RW}}_{t+1}(\tau)\gets y_{t+1}(\tau)-y_t(\tau)$
\For{each specification $s$, maturity $\tau$}
  \State $d_t\gets e^{s}_{t+1}(\tau)^2-e^{\mathrm{RW}}_{t+1}(\tau)^2$;\quad
         $\bar d\gets\tfrac1n\sum_t d_t$
  \State $\hat\omega\gets$ Newey--West long-run variance of $\{d_t\}$ (lag $h-1$)
  \State $\mathrm{DM}^{s}(\tau)\gets\big(\bar d/\sqrt{\hat\omega/n}\big)\cdot
         \sqrt{(n+1-2h+h(h-1)/n)/n}$
         \Comment{Harvey--Leybourne--Newbold correction}
  \State $p^{s}(\tau)\gets 2\,F_{t_{n-1}}\!\big(-|\mathrm{DM}^{s}(\tau)|\big)$
\EndFor
\For{each maturity $\tau$}
  \State report $\operatorname{median}_s\mathrm{DM}^{s}(\tau)$ and
         $[\min_s,\max_s]\,\mathrm{DM}^{s}(\tau)$
  \State significant at $\alpha$ iff $p^{s}(\tau)<\alpha$ for all $s$
         \Comment{a negative starred value rejects in the forecaster's favor}
\EndFor
\end{algorithmic}
\end{algorithm}

\bibliography{references}

@article{nelson1987parsimonious,
  author  = {Nelson, Charles R. and Siegel, Andrew F.},
  title   = {Parsimonious Modeling of Yield Curves},
  journal = {Journal of Business},
  volume  = {60},
  number  = {4},
  pages   = {473--489},
  year    = {1987}
}

@article{diebold2006forecasting,
  author  = {Diebold, Francis X. and Li, Canlin},
  title   = {Forecasting the Term Structure of Government Bond Yields},
  journal = {Journal of Econometrics},
  volume  = {130},
  number  = {2},
  pages   = {337--364},
  year    = {2006}
}

@article{fernandes2019dynamic,
  author  = {Fernandes, Marcelo and Vieira, Fausto},
  title   = {A Dynamic {N}elson--{S}iegel Model with Forward-Looking Macroeconomic Factors for the Yield Curve in the {US}},
  journal = {Journal of Economic Dynamics and Control},
  volume  = {106},
  pages   = {103720},
  year    = {2019},
  doi     = {10.1016/j.jedc.2019.103720}
}

@article{breiman2001random,
  author  = {Breiman, Leo},
  title   = {Random Forests},
  journal = {Machine Learning},
  volume  = {45},
  number  = {1},
  pages   = {5--32},
  year    = {2001}
}

@techreport{cartea2025limited,
  author      = {Cartea, {\'A}lvaro and Jin, Qi and Shi, Yuantao},
  title       = {The Limited Virtue of Complexity in a Noisy World},
  institution = {University of Oxford},
  year        = {2025},
  note        = {Available at SSRN: https://ssrn.com/abstract=5202064}
}

@article{caldeira2016forecast,
  author  = {Caldeira, Jo{\~a}o F. and Moura, Guilherme V. and Santos, Andr{\'e} A. P.},
  title   = {Predicting the Yield Curve Using Forecast Combinations},
  journal = {Computational Statistics \& Data Analysis},
  volume  = {100},
  pages   = {79--98},
  year    = {2016},
  doi     = {10.1016/j.csda.2014.05.008}
}

@article{granger1984improved,
  author  = {Granger, Clive W. J. and Ramanathan, Ramu},
  title   = {Improved Methods of Combining Forecasts},
  journal = {Journal of Forecasting},
  volume  = {3},
  number  = {2},
  pages   = {197--204},
  year    = {1984}
}

@article{yang2004combining,
  author  = {Yang, Yuhong},
  title   = {Combining Forecasting Procedures: Some Theoretical Results},
  journal = {Econometric Theory},
  volume  = {20},
  number  = {1},
  pages   = {176--222},
  year    = {2004}
}

@article{jiang2025robust,
  author  = {Jiang, Binyan and Lv, Jing and Li, Jialiang and Cheng, Ming-Yen},
  title   = {Robust Model Averaging Prediction of Longitudinal Response with Ultrahigh-Dimensional Covariates},
  journal = {Journal of the Royal Statistical Society: Series B (Statistical Methodology)},
  volume  = {87},
  number  = {2},
  pages   = {337--361},
  year    = {2025}
}

@article{delage2010distributionally,
  author  = {Delage, Erick and Ye, Yinyu},
  title   = {Distributionally Robust Optimization under Moment Uncertainty with Application to Data-Driven Problems},
  journal = {Operations Research},
  volume  = {58},
  number  = {3},
  pages   = {595--612},
  year    = {2010}
}

@article{MohajerinEsfahani2018,
  author  = {Mohajerin Esfahani, Peyman and Kuhn, Daniel},
  title   = {Data-Driven Distributionally Robust Optimization Using
             the {W}asserstein Metric: Performance Guarantees and
             Tractable Reformulations},
  journal = {Mathematical Programming},
  volume  = {171},
  number  = {1--2},
  pages   = {115--166},
  year    = {2018}
}

@incollection{kuhn2019wasserstein,
  author    = {Kuhn, Daniel and Mohajerin Esfahani, Peyman and
               Nguyen, Viet Anh and Shafieezadeh-Abadeh, Soroosh},
  title     = {Wasserstein Distributionally Robust Optimization:
               Theory and Applications in Machine Learning},
  booktitle = {Operations Research \& Management Science in the Age of Analytics},
  publisher = {INFORMS},
  pages     = {130--166},
  year      = {2019},
}

@inproceedings{nguyen2020distributionally,
  author    = {Nguyen, Viet Anh and Zhang, Fan and Blanchet, Jose and Delage, Erick and Ye, Yinyu},
  booktitle = {Advances in Neural Information Processing Systems},
  volume    = {33},
  pages     = {15232--15242},
  title     = {Distributionally Robust Local Non-Parametric Conditional Estimation},
  year      = {2020}
}

@article{KannanBayraksanLuedtke2024,
  author  = {Kannan, Rohit and Bayraksan, G{\"u}zin and Luedtke, James R.},
  title   = {Residuals-Based Distributionally Robust Optimization with Covariate Information},
  journal = {Mathematical Programming},
  volume  = {207},
  pages   = {369--425},
  year    = {2024},
  doi     = {10.1007/s10107-023-02014-7}
}

@article{NguyenZhangWangBlanchetDelageYe2024,
  author  = {Nguyen, Viet Anh and Zhang, Fan and Wang, Shanshan and Blanchet, Jose and Delage, Erick and Ye, Yinyu},
  title   = {Robustifying Conditional Portfolio Decisions via Optimal Transport},
  journal = {Operations Research},
  volume  = {73},
  number  = {5},
  pages   = {2801--2829},
  year    = {2025}
}

@article{BaiNg2002,
  author  = {Bai, Jushan and Ng, Serena},
  title   = {Determining the Number of Factors in Approximate
             Factor Models},
  journal = {Econometrica},
  volume  = {70},
  number  = {1},
  pages   = {191--221},
  year    = {2002}
}

@article{Onatski2012,
  author  = {Onatski, Alexei},
  title   = {Asymptotics of the principal components estimator of large factor models with weakly influential factors},
  journal = {Journal of Econometrics},
  volume  = {168},
  number  = {2},
  pages   = {244--258},
  year    = {2012},
  doi     = {10.1016/j.jeconom.2012.01.034}
}

@article{LettauPelger2020,
  author  = {Lettau, Martin and Pelger, Markus},
  title   = {Factors That Fit the Time Series and Cross-Section of Stock Returns},
  journal = {Review of Financial Studies},
  year    = {2020},
  volume  = {33},
  number  = {5},
  pages   = {2274--2325}
}

@article{Tibshirani1996,
  author  = {Tibshirani, Robert},
  title   = {Regression Shrinkage and Selection via the Lasso},
  journal = {Journal of the Royal Statistical Society,
             Series B},
  volume  = {58},
  number  = {1},
  pages   = {267--288},
  year    = {1996}
}

@article{Wainwright2009,
  author  = {Wainwright, Martin J.},
  title   = {Sharp Thresholds for High-Dimensional and Noisy Sparsity
             Recovery Using $\ell_1$-Constrained Quadratic Programming
             ({L}asso)},
  journal = {IEEE Transactions on Information Theory},
  volume  = {55},
  number  = {5},
  pages   = {2183--2202},
  year    = {2009}
}

@article{LedoitWolf2004,
  author  = {Ledoit, Olivier and Wolf, Michael},
  title   = {Honey, {I} Shrunk the Sample Covariance Matrix},
  journal = {The Journal of Portfolio Management},
  volume  = {30},
  number  = {4},
  pages   = {110--119},
  year    = {2004},
  doi     = {10.3905/jpm.2004.110}
}

@article{RockafellarUryasev2000,
  author  = {Rockafellar, R. Tyrrell and Uryasev, Stanislav},
  title   = {Optimization of Conditional Value-at-Risk},
  journal = {Journal of Risk},
  volume  = {2},
  pages   = {21--41},
  year    = {2000}
}

@article{dickey1979distribution,
  author  = {Dickey, David A. and Fuller, Wayne A.},
  title   = {Distribution of the Estimators for Autoregressive Time Series with a Unit Root},
  journal = {Journal of the American Statistical Association},
  volume  = {74},
  number  = {366},
  pages   = {427--431},
  year    = {1979}
}

@article{fan2020factor,
  author  = {Fan, Jianqing and Ke, Yuan and Wang, Kaizheng},
  title   = {Factor-Adjusted Regularized Model Selection},
  journal = {Journal of Econometrics},
  year    = {2020},
  volume  = {216},
  number  = {1},
  pages   = {71--85}
}

@article{fan2001variable,
  author  = {Fan, Jianqing and Li, Runze},
  title   = {Variable Selection via Nonconcave Penalized Likelihood and Its Oracle Properties},
  journal = {Journal of the American Statistical Association},
  year    = {2001},
  volume  = {96},
  number  = {456},
  pages   = {1348--1360}
}

@article{ahn2013eigenvalue,
  author  = {Ahn, Seung C. and Horenstein, Alex R.},
  title   = {Eigenvalue Ratio Test for the Number of Factors},
  journal = {Econometrica},
  year    = {2013},
  volume  = {81},
  number  = {3},
  pages   = {1203--1227}
}

@article{huber1964robust,
  author  = {Huber, Peter J.},
  title   = {Robust Estimation of a Location Parameter},
  journal = {The Annals of Mathematical Statistics},
  year    = {1964},
  volume  = {35},
  number  = {1},
  pages   = {73--101}
}

@article{diebold1995comparing,
  author  = {Diebold, Francis X. and Mariano, Roberto S.},
  title   = {Comparing Predictive Accuracy},
  journal = {Journal of Business \& Economic Statistics},
  year    = {1995},
  volume  = {13},
  number  = {3},
  pages   = {253--263}
}

@article{baker2016measuring,
  author  = {Baker, Scott R. and Bloom, Nicholas and Davis, Steven J.},
  title   = {Measuring Economic Policy Uncertainty},
  journal = {The Quarterly Journal of Economics},
  year    = {2016},
  volume  = {131},
  number  = {4},
  pages   = {1593--1636}
}

@article{BatesGranger1969,
  author  = {Bates, J. M. and Granger, C. W. J.},
  title   = {The Combination of Forecasts},
  journal = {Operational Research Quarterly},
  year    = {1969},
  volume  = {20},
  number  = {4},
  pages   = {451--468},
  doi     = {10.1057/jors.1969.103}
}

@article{harvey1997testing,
  author  = {Harvey, David and Leybourne, Stephen and Newbold, Paul},
  title   = {Testing the Equality of Prediction Mean Squared Errors},
  journal = {International Journal of Forecasting},
  year    = {1997},
  volume  = {13},
  number  = {2},
  pages   = {281--291}
}

@article{duffee2002term,
  author  = {Duffee, Gregory R.},
  title   = {Term Premia and Interest Rate Forecasts in Affine Models},
  journal = {Journal of Finance},
  volume  = {57},
  number  = {1},
  pages   = {405--443},
  year    = {2002}
}

\end{document}